\documentclass[letter]{aa} % for the letters
\usepackage{natbib}
\usepackage{multirow}
\usepackage{graphicx}
\usepackage{txfonts}
\newcommand{\jmst}{J.~Mol.~Struct.}   % Journal of Molecular Structure

\newcommand{\kms}{km s$^{-1}$}
\newcommand{\ten}{10$^{10}$\,cm$^{-2}$}
\newcommand{\once}{10$^{11}$\,cm$^{-2}$}
\newcommand{\doce}{10$^{12}$\,cm$^{-2}$}
\newcommand{\trece}{10$^{13}$\,cm$^{-2}$}

\begin{document}

\title{TMC-1, the starless core sulfur factory: Discovery of NCS, HCCS, H$_2$CCS, 
H$_2$CCCS, and C$_4$S and detection of C$_5$S
%  TMC-1, the starless core that turns to be a sulfur factory: detection of C$_5$S and
%  discovery of NCS, HCCS, H$_2$CCS, H$_2$CCCS, and C$_4$S 
\thanks{Based on observations carried out
with the Yebes 40m telescope (projects 19A003,
20A014, and 20D023) and the Institut de Radioastronomie Millim\'etrique (IRAM) 30m telescope. The 40m
radiotelescope at Yebes Observatory is operated by the Spanish Geographic Institute
(IGN, Ministerio de Transportes, Movilidad y Agenda Urbana). IRAM is supported by INSU/CNRS
(France), MPG (Germany), and IGN (Spain).}}

\author{
J.~Cernicharo\inst{1},
C.~Cabezas\inst{1},
M.~Ag\'undez\inst{1},
B.~Tercero\inst{2,3}, 
J.~R.~Pardo\inst{1},
N.~Marcelino\inst{1},
J.D.~Gallego\inst{3},
F.~Tercero\inst{3}, 
J.A.~L\'opez-P\'erez\inst{3},
and P.~de~Vicente\inst{3},
}

\institute{Grupo de Astrof\'isica Molecular, Instituto de F\'isica Fundamental (IFF-CSIC), C/ Serrano 121, 28006 Madrid, Spain.\\
\email: jose.cernicharo@csic.es
\and Observatorio Astron\'omico Nacional (IGN), C/ Alfonso XII, 3, 28014, Madrid, Spain.
\and Centro de Desarrollos Tecnol\'ogicos, Observatorio de Yebes (IGN), 19141 Yebes, Guadalajara, Spain.
}

\date{Received; accepted}

\abstract{We report the detection  of the sulfur-bearing species NCS, HCCS, H$_2$CCS, 
H$_2$CCCS, and C$_4$S for the first time in space. 
These molecules were found towards TMC-1 through the observation of several lines for each species.
We also report the detection of C$_5$S for the first time in a cold cloud through the observation of five lines in the 
31-50 GHz range. 
The derived column densities are 
$N$(NCS) = (7.8$\pm$0.6)$\times$\once,
$N$(HCCS) = (6.8$\pm$0.6)$\times$\once,
$N$(H$_2$CCS) = (7.8$\pm$0.8)$\times$\once,
$N$(H$_2$CCCS) = (3.7$\pm$0.4)$\times$\once,
$N$(C$_4$S) = (3.8$\pm$0.4)$\times$10$^{10}$ cm$^{-2}$, and
$N$(C$_5$S) = (5.0$\pm$1.0)$\times$10$^{10}$ cm$^{-2}$. The observed abundance ratio between C$_3$S and C$_4$S is 340, 
that is to say a factor of approximately one hundred larger than the corresponding 
value for CCS and C$_3$S. The observational results are compared with 
a state-of-the-art chemical model, which is only partially successful in reproducing the 
observed abundances. These detections underline the need to improve chemical networks dealing with S-bearing species.}

\keywords{ Astrochemistry
---  ISM: molecules
---  ISM: individual (TMC-1)
---  line: identification
---  molecular data}

\titlerunning{Sulfur-bearing species in TMC-1}
\authorrunning{Cernicharo et al.}

\maketitle

\section{Introduction}
The cold dark core TMC-1 presents an interesting carbon-rich chemistry that leads to
the formation of long neutral carbon-chain radicals and their anions, as well as cyanopolyynes
(see \citealt{Cernicharo2020a,Cernicharo2020b} and references therein). The carbon chains 
CCS and CCCS are
particularly abundant in this cloud \citep{Saito1987,Yamamoto1987} and also exist in the
envelopes of carbon-rich circumstellar envelopes \citep{Cernicharo1987}. TMC-1 is also
peculiar due to the presence of protonated species of abundant large carbon chains such
as HC$_3$O$^+$ \citep{Cernicharo2020c}, HC$_5$NH$^+$ \citep{Marcelino2020},
HC$_3$S$^+$ \citep{Cernicharo2021a}, and CH$_3$CO$^+$ \citep{Cernicharo2021b}. 
The number of sulfur-bearing species detected to date in TMC-1 is small 
compared to oxygen- and nitrogen-bearing species (see, e.g. \citealt{McGuire2019}). In fact, the chemistry of sulfur-bearing molecules is strongly
dependent on the depletion of sulfur \citep{Vidal2017}. Many reactions involving
S$^+$ with neutrals as well as radicals with S and CS have to be studied to achieve a
better chemical modelling of sulfur-bearing species \citep{Petrie1996,Bulut2021}. Nevertheless, 
the main input to understand these chemical processes is to unveil new sulfur-bearing
species in the interstellar medium as well as to understand their formation paths and their
role in the chemistry of sulfur.

In this letter we report the discovery, for the first time in space, of the following five new
sulfur-bearing species:  
NCS, HCCS, H$_2$CCS, H$_2$CCCS, and C$_4$S. 
The detection of C$_5$S in a cold dark cloud
is also reported for the first time. A detailed observational study of the most 
relevant S-bearing species in this
cloud is accomplished. We discuss these results in the context of state-of-the-art chemical models. 

\vspace{-0.2cm}
\section{Observations}
New receivers, built
within the Nanocosmos
project\footnote{\texttt{https://nanocosmos.iff.csic.es/}}, and installed at the Yebes 40m radio telescope were used
for the observations of TMC-1. 
The receivers and the spectrometers have been described by \citet{Tercero2021}. 
The observations needed to complete the Q-band line survey towards TMC-1
($\alpha_{J2000}=4^{\rm h} 41^{\rm  m} 41.9^{\rm s}$ and $\delta_{J2000}=+25^\circ 41' 27.0''$)
were performed in several sessions during December 2019 and January 2021.
All data were analysed using the GILDAS package\footnote{\texttt{http://www.iram.fr/IRAMFR/GILDAS}}.
The observing procedure
has been previously described (see, e.g. \citealt{Cernicharo2021a,Cernicharo2021b}). 
The IRAM 30m data come from a line survey performed towards TMC-1 and B1 and the observations
have been described by \citet{Marcelino2007} and \cite{Cernicharo2013}.

The intensity scale, antenna temperature ($T_A^*$) was calibrated using two absorbers
at different temperatures and an atmospheric transmission model (ATM; \citealt{Cernicharo1985, Pardo2001}). Calibration
uncertainties have been adopted to be 10~\%. The nominal spectral resolution of 38.15 kHz was used for the final
spectra. 

\section{Results and discussion}
\label{results}
Line identification in our TMC-1 survey has been performed using the MADEX catalogue \citep{Cernicharo2012},
the Cologne Database of Molecular Spectroscopy catalogue \citep{Muller2005}, and the JPL catalogue \citep{Pickett1998}.
A description of the methods used to fit the data and to derive
column densities is provided in Appendix \ref{appendix_cd}.

\begin{figure}[]
\centering
\includegraphics[width=0.95\columnwidth,angle=0]{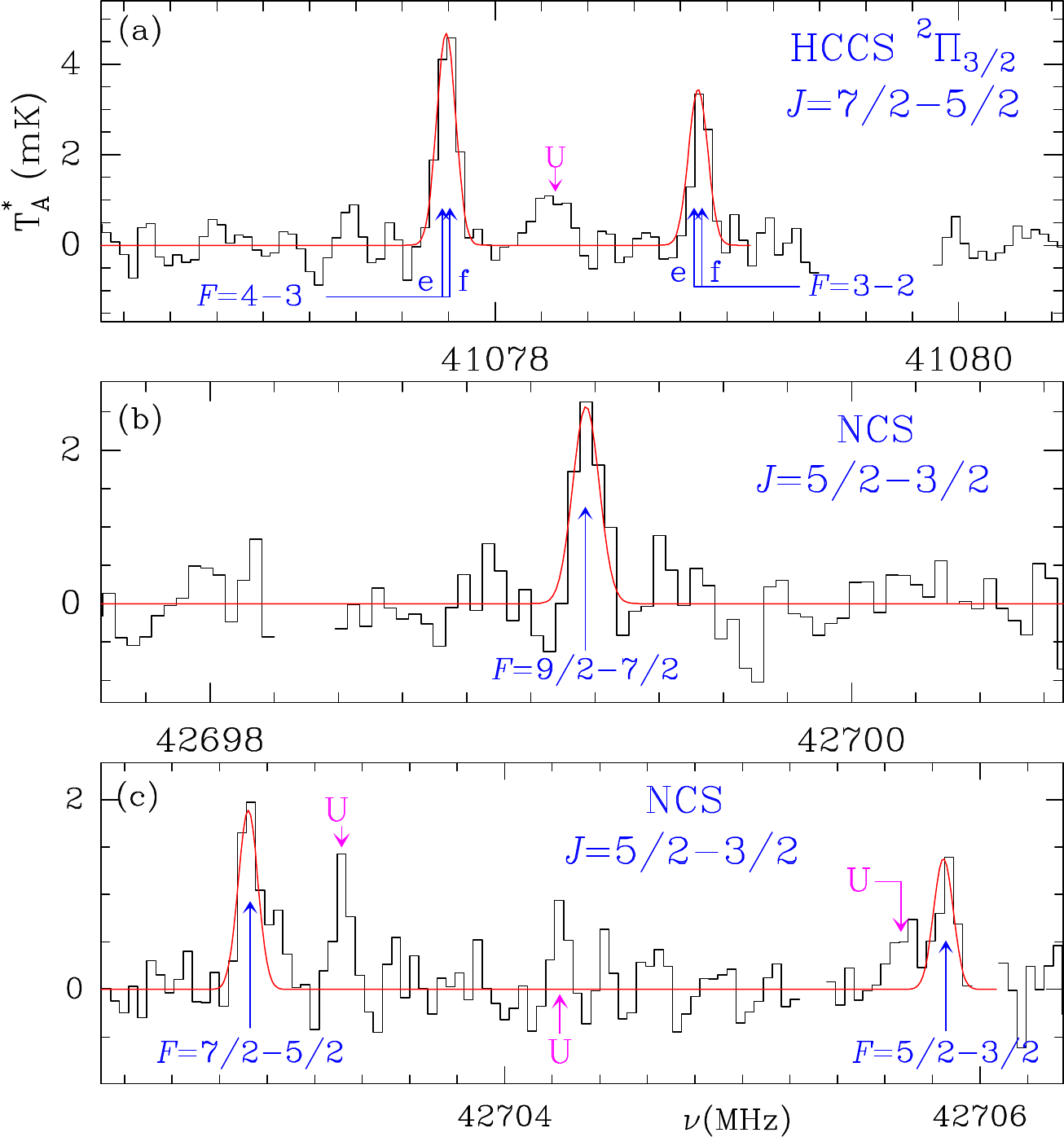}
\caption{Observed lines of HCCS (panel $a$) and NCS (panels $b$ and $c$) towards TMC-1.
The abscissa corresponds to the rest frequency assuming a local standard of rest velocity of the source of
5.83 km s$^{-1}$ (see text). Blanked channels correspond to negative features produced in
the frequency switching data folding.
%Frequencies and intensities for the observed lines are given in Table \ref{tab_s_bearing}.
The ordinate is the antenna temperature corrected for atmospheric and telescope losses in 
milliKelvin. 
Spectral resolution is 38.15 kHz. The red lines show the synthetic spectrum of HCCS and NCS for a rotational
temperature of 5 K, a linewidth of 0.6 km s$^{-1}$, and a column density of 6.8$\times$10$^{11}$ cm$^{-2}$
and 7.8$\times$10$^{11}$ cm$^{-2}$, respectively.}
\label{fig_hccs}
\end{figure}

\subsection{HCCS and NCS}
Among the unidentified spectral features, we have found a couple separated by 1 MHz that perfectly match the frequencies of the strongest hyperfine components of the $J$=7/2-5/2 transition of HCCS ($^2\Pi_{3/2}$). This species was observed in the laboratory by \citet{Kim2002} and the
prediction of its rotational spectrum is available in the CDMS \citep{Muller2005} and MADEX \citep{Cernicharo2012} 
catalogues. Taking into account the perfect match in frequencies, the narrow linewidth of the emission features in
this source, and the perfect match in the relative intensities of the two hyperfine components, 
the possibility of a fortuitous coincidence is very low. Hence, we conclude that these two lines arise from
HCCS. The observed lines are shown in Fig.\,\ref{fig_hccs} and the derived line parameters are
given in Table \ref{tab_s_bearing}. The synthetic spectrum on Fig.\,\ref{fig_hccs} corresponds to
T$_r$=5 K and $N$(HCCS) = 6.8$\times$\once.
We searched for HCCCS but only a 3$\sigma$ upper limit to its 
column density of 2.4$\times$\once\, can be given (see Table \ref{column_densities}).

\begin{figure}[]
\centering
\includegraphics[width=0.91\columnwidth,angle=0]{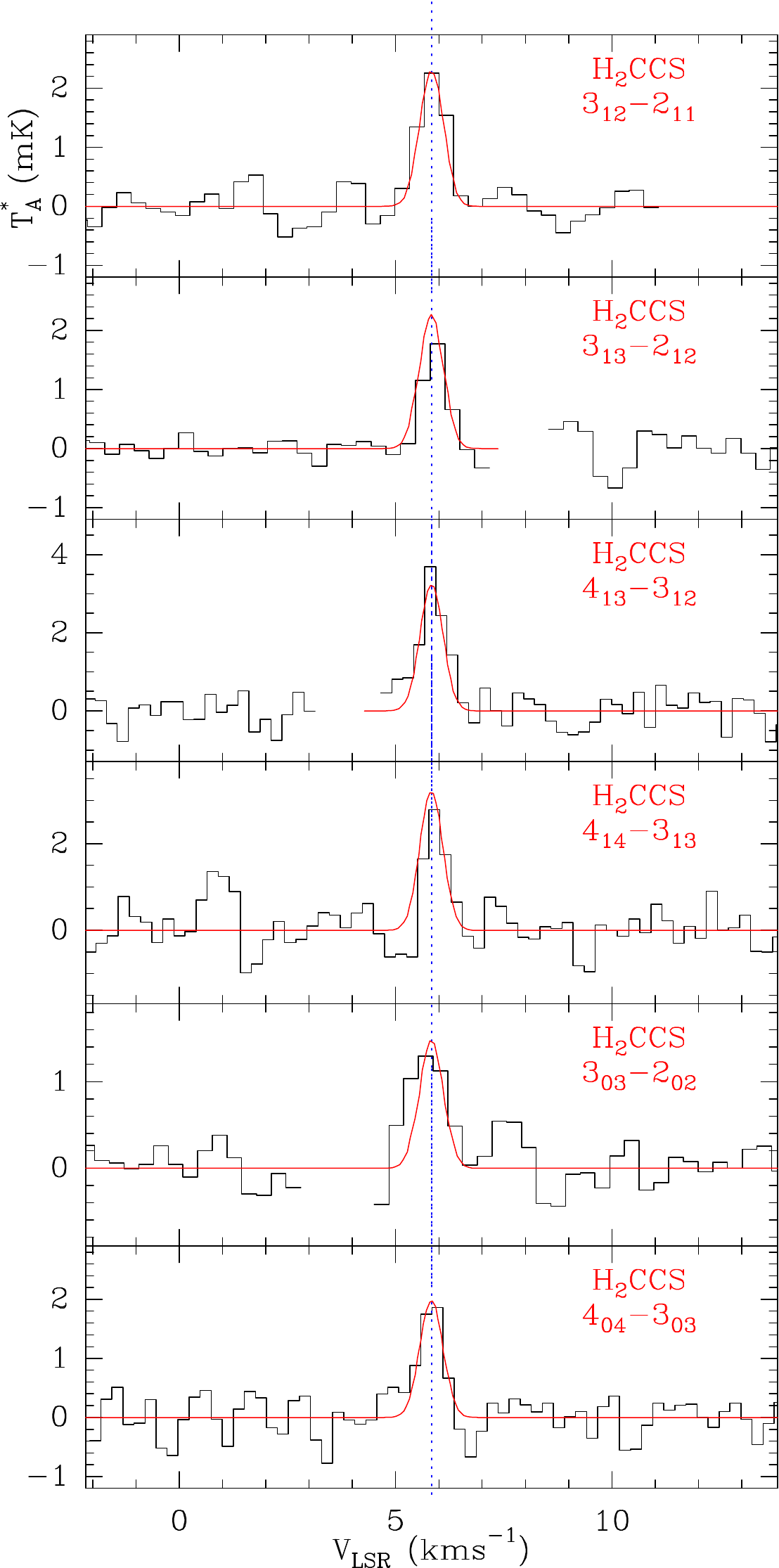}
\caption{Observed transitions of H$_2$CCS towards TMC-1.
The abscissa corresponds to the velocity in km\,s$^{-1}$ of the source with respect to the
local standard of rest. Line parameters are given in Table \ref{tab_s_bearing}.
The ordinate is the antenna temperature corrected for atmospheric and telescope losses in milliKelvin. 
Blanked channels correspond to negative features produced by the folding of the frequency
switching observations. The vertical blue dashed line indicates the v$_{LSR}$ of the
cloud (5.83 km\,s$^{-1}$). The red line spectra correspond to the synthetic model
spectrum for each line adopting T$_r$=7 K, $N$($o$-H$_2$CCS) = 6.0$\times$\once, and
$N$($p$-H$_2$CCS) = 1.8$\times$\once\, (see text).}
\label{fig_h2ccs}
\end{figure}

Thiocyanogen, NCS, 
has been observed in the laboratory \citep{Amano1991,McCarthy2003,Maeda2007}, but never detected in 
space. Only one rotational transition lies within the Q-band, the $J$=5/2-3/2 transition of NCS 
in its $^2\Pi_{3/2}$ ladder. 
Figure \ref{fig_hccs} shows the three hyperfine components of this transition observed in TMC-1. The match
between observations and the synthetic spectrum, corresponding to T$_r$=5 K and $N$(NCS) = 7.8$\times$\once\,, 
is remarkably good, ensuring the discovery of this sulfur compound in TMC-1. The oxygen analogue of thiocyanogen,
the isocyanate radical (NCO), was detected in cold core L483 by \citet{Marcelino2018}. Unfortunately,
NCO does not have lines in the 31-50 GHz range.

\subsection{H$_2$CCS and H$_2$CCCS}

Prompted by the detection of HCCS, we searched for other sulfur-bearing species in our survey.
\citet{McGuire2019} searched for H$_2$CCS, thioketene,  in TMC-1. They obtained an upper 
limit to its column density of 5.5$\times$\doce. Unfortunately they used a transition with an upper level energy
around 40 K, which is not the best for the physical conditions of TMC-1. In this work we report the 
discovery of thioketene in space through observations of transitions with smaller upper level energies.

\begin{figure}[]
\centering
\includegraphics[width=0.85\columnwidth,angle=0]{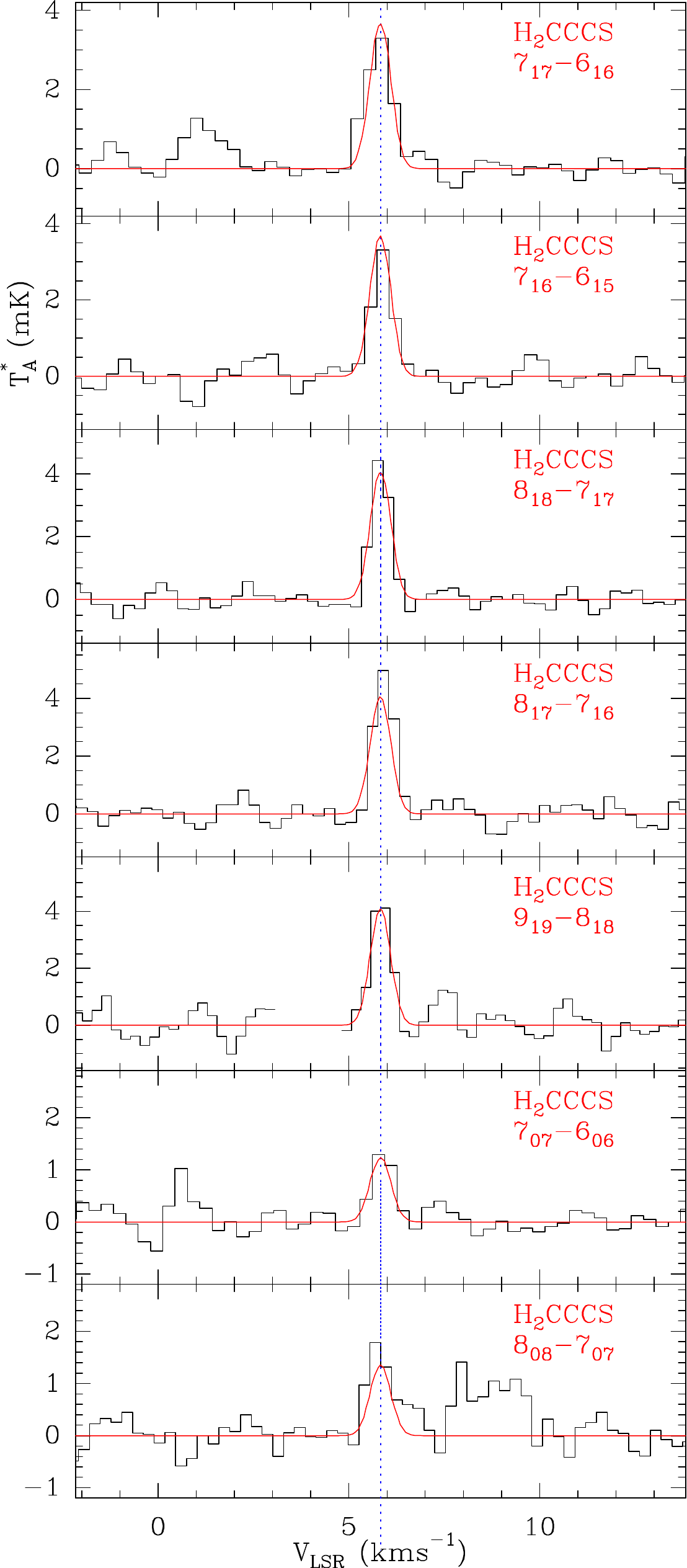}
\caption{Selected transitions of H$_2$CCCS towards TMC-1.
The abscissa corresponds to the velocity in km\,s$^{-1}$ of the source with respect to the
local standard of rest. Line parameters are given in Table \ref{tab_s_bearing}.
The ordinate is the antenna temperature corrected for atmospheric and telescope losses in milliKelvin. 
Blanked channels correspond to negative features produced by the folding of the frequency
switching observations. The vertical blue dashed line indicates the v$_{LSR}$ of the
cloud (5.83 km\,s$^{-1}$). The red line spectra correspond to the synthetic model
of the emission obtained for T$_r$=10 K, $N$($o$-H$_2$CCS) = 3.0$\times$\once, and
$N$($p$-H$_2$CCS) = 6.5$\times$\ten.}
\label{fig_h2cccs}
\end{figure}

Spectroscopic laboratory
data of thioketene \citep{Georgiou1979,Winnewiser1980,Naughton1996} and H$_2$CCCS \citep{Brown1988} 
were used to predict the frequencies of these species and implement them in the MADEX code.
The dipole moments adopted for H$_2$CCS and H$_2$CCCS are 1.02 D \citep{Georgiou1979} and
2.06 D \citep{Brown1988}, respectively.
We detected the four ortho and two para transitions of H$_2$CCS expected in our data, as shown
in Fig. \ref{fig_h2ccs}. 
For H$_2$CCCS (propadienthione), six ortho and three para transitions are detected
(see seven of them in Fig. \ref{fig_h2cccs}). 
The derived line parameters for both species are given in Table \ref{tab_s_bearing}. 

An analysis of the data through a standard rotation diagram provides rotational temperatures 
of 7$\pm1$ K and 10$\pm$1 K for H$_2$CCS and H$_2$CCCS, respectively. For H$_2$CCS,
we derived $N$(ortho) = (6.0$\pm$0.6)$\times$\once\, and 
$N$(para) = (1.8$\pm$0.2)$\times$\once. Hence, the
ortho/para ratio for this species is 3.3$\pm$0.7. For H$_2$CCCS, we
derived $N$(ortho) = (3.0$\pm$0.3)$\times$\once\, and $N$(para) = (6.5$\pm$0.5)$\times$\ten, respectively.
The ortho/para abundance ratio derived for this species is 4.6$\pm$0.8. These ortho/para ratios are compatible with 
the 3/1 value expected from the statitiscal spin degeneracies. They imply that non-significant enrichment of the para species is
produced through reactions of H$_2$CCS and H$_2$CCCS with H$_3^+$, or other protonated
molecular cations.
The synthetic spectrum computed with these
parameters for both species is shown in red in the different panels of 
Figures \ref{fig_h2ccs} and \ref{fig_h2cccs}.
The derived abundance ratio between H$_2$CCS and H$_2$CCCS is $\simeq$2, which
is very different from the H$_2$CCO/H$_2$CCCO abundance ratio of $>$\,130 derived
in the same source by \citet{Cernicharo2020c}, but very similar to the C$_2$S/C$_3$S abundance
ratio of $\sim$3 derived here (see Table \ref{column_densities}).

\subsection{C$_4$S and C$_5$S}
Taking the large column density derived for CCS and C$_3$S into account \citep{Cernicharo2021a}, the next
member of this family, C$_4$S, 
is a potential candidate to be present in TMC-1. 
The spectroscopic laboratory data used to predict the spectrum of C$_4$S are from 
\citet{Hirahara1993} and \citet{Gordon2001}. A 
dipole moment of 4.03 D was computed through ab initio calculations by \citet{Pascoli1998}
and \citet{Lee1997}.
Nineteen transitions of this species have frequencies within the range of our line survey. 
The line by line
search through our data provides a clear detection for the $N_u$=10 up to 13, $J_u$=$N_u$+1 
transitions (see Fig. \ref{fig_c4s}).
A rotational temperature of 7$\pm$1 K was derived from these lines. The model fitting method (see Appendix \ref{appendix_cd}) was
used to derive a column density of (3.8$\pm$0.4)$\times$\ten. The stacked spectrum obtained for
these lines and for the weaker $J_u$=$N$ and $J_u$=$N$-1 transitions are
shown in panels (b) and (c) of Fig. \ref{fig_c4s}, respectively.

The next member
of the C$_n$S family, C$_5$S, was tentatively detected towards the carbon-rich star CW Leo by \citet{Bell1993}
and confirmed by \citet{Agundez2014}. Laboratory spectroscopy from \citet{Gordon2001} 
and a dipole moment of 4.65 D \citep{Pascoli1998} have been adopted. Five lines from $J_u$=17 up to 21
have been detected (see Fig. \ref{fig_c5s}). 
The model fitting provides a rotational temperature of 7$\pm$2\,K and a column density of  
(5.0$\pm$1.0)$\times$10$^{10}$ cm$^{-2}$. The abundance ratio between C$_4$S and C$_5$S is of the
order of unity, and the abundance ratio C$_2$S/C$_3$S/C$_4$S/C$_5$S is 5500/1300/3.8/5.0. The change in
abundance for C$_4$S and C$_5$S relative to C$_2$S is of three orders of magnitude. This result
is very different than the one obtained by \citet{Agundez2014} for the carbon-rich star 
IRC+10216. In this object, the derived
C$_3$S/C$_5$S ratio is $\sim$1-10 (depending on the assumed rotational temperature),
versus $\sim$260 in TMC-1. However, the C$_2$S/C$_3$S abundance ratio is the same for both
sources, that is to say $\sim$3. The radical C$_4$S has not been detected yet in IRC+10216 \citep{Agundez2014}.
S-bearing carbon chains do not follow the
smooth decrease in abundance observed in cold dark clouds and circumstellar envelopes for other
carbon chains such as cyanopolyynes (HC$_{2n+1}$N; a factor 3-5 between members of this molecular
species).

\begin{table}
\caption{Column densities of sulfur-bearing species in TMC-1.}
\label{column_densities}
\centering
\small
\begin{tabular}{{lccc}}
\hline
Molecule                          & T$_{rot}$ (K) &        $N_{\rm obs}$ (cm$^{-2}$) & $N_{\rm calc}$ (cm$^{-2}$)\,$^e$ \\
\hline               
CS$^{a*}$            & 10.0 & (3.50$\pm$0.40)$\times$10$^{14}$ & $3.1 \times 10^{14}$   \\
C$^{34}$S$^*$        & 10.0 & (1.45$\pm$0.10)$\times$10$^{13}$  \\
$^{13}$C$^{34}$S$^*$ & 10.0 & (1.45$\pm$0.20)$\times$10$^{11}$  \\
HCS                  &  5.0$^b$ & (5.50$\pm$0.50)$\times$10$^{12}$ & $5.1 \times 10^{10}$ \\ 
HSC                  &  5.0$^b$ & (1.30$\pm$0.20)$\times$10$^{11}$  \\
NCS                  &  5.0$^b$ & (7.80$\pm$0.60)$\times$10$^{11}$ & $2.5 \times 10^{10}$ \\ 
HCCS                 &  5.0$^b$ & (6.80$\pm$0.60)$\times$10$^{11}$ & $3.1 \times 10^{10}$ \\ 
H$_2$CS$^c$          & 10.0 & (4.70$\pm$0.40)$\times$10$^{13}$     & $9.5 \times 10^{12}$ \\
H$_2$C$^{34}$S$^c$   & 10.0 & (1.80$\pm$0.18)$\times$10$^{12}$  \\
$o$-H$_2$CCS         &  7.0$\pm$1.0 & (6.00$\pm$0.60)$\times$10$^{11}$      & \multirow{2}{*}{$\Big\} 1.8 \times 10^{12}$} \\ 
$p$-H$_2$CCS         &  7.0$\pm$1.0 & (1.80$\pm$0.20)$\times$10$^{11}$     & \\ 
$o$-H$_2$CCCS        & 10.0$\pm$1.0 & (3.00$\pm$0.30)$\times$10$^{11}$ & \multirow{2}{*}{$\Big\} 7.8 \times 10^{10}$} \\
$p$-H$_2$CCCS        & 10.0$\pm$1.0 & (6.50$\pm$0.60)$\times$10$^{10}$ & \\ 
CCS$^*$              &  5.1$\pm$0.2 & (5.50$\pm$0.65)$\times$10$^{13}$           & $3.8 \times 10^{11}$\\
CC$^{34}$S$^*$       &  3.6$\pm$0.4 & (5.00$\pm$0.50)$\times$10$^{12}$      \\
CCCS$^*$             &  5.8$\pm$0.2 & (1.30$\pm$0.13)$\times$10$^{13}$          & $1.1 \times 10^{13}$\\
CCC$^{34}$S$^*$      &  6.7$\pm$0.2 & (5.30$\pm$0.50)$\times$10$^{11}$  \\
C$_4$S               & 7.0$\pm$1.0 & (3.80$\pm$0.50)$\times$10$^{10}$            & $1.6 \times 10^{10}$ \\ 
C$_5$S               & 7.0$\pm$1.0 & (5.00$\pm$1.00)$\times$10$^{10}$            & $1.9 \times 10^{11}$ \\ 
HCS$^{+*}$           & 10.0 & (1.00$\pm$0.10)$\times$10$^{13}$                        & $1.7 \times 10^{11}$\\
HC$^{34}$S$^{+*}$    & 10.0 & (7.10$\pm$0.70)$\times$10$^{11}$   \\
HC$_3$S$^+$$^*$      & 10.0$\pm$2.0 & (4.00$\pm$1.50)$\times$10$^{11}$   & $2.8 \times 10^{11}$ \\
HSCN                 &  5.0 & (5.80$\pm$0.60)$\times$10$^{11}$                            & $3.7 \times 10^{9}$ \\ 
HNCS                 &  5.0 & (3.80$\pm$0.40)$\times$10$^{11}$                            & $1.2 \times 10^{10}$  \\
HC$_3$S$^d$          &  5.0 & $\leq$2.40$\times$10$^{11}$                               & $1.5 \times 10^{11}$ \\ 
HCNS$^d$             & 10.0 & $\leq$6.00$\times$10$^{10}$                                & $3.9 \times 10^{6}$  \\
HSNC$^d$             & 10.0 & $\leq$2.00$\times$10$^{10}$        \\
HSCS$^+$$^d$         &  5.0 & $\leq$4.00$\times$10$^{12}$        \\ 
HCCSH$^d$            &  7.0 & $\leq$3.00$\times$10$^{12}$        \\ 
HCCCSH$^d$           &  7.0 & $\leq$2.40$\times$10$^{11}$        \\ 
\hline
\end{tabular}
\tablefoot{\\
 Entries in bold face correspond to the molecular species detected for the first time
 in space, or in TMC-1 (C$_5$S). \\
        \tablefoottext{*}{Data from \citet{Cernicharo2021a}.} \\
        \tablefoottext{a}{Derived from C$^{34}$S and the C$_3$S/C$_3$$^{34}$S abundance ratio.}\\
        \tablefoottext{b}{Rotational temperature assumed identical to that of CCS.}\\
        \tablefoottext{c}{Column density refers to the ortho or para species. Assuming an ortho/para ratio of 3,
    the total column density can be estimated by multiplying the para value by a factor of 4.}\\
        \tablefoottext{d}{Upper limits correspond to 3$\sigma$ values derived assuming the indicated rotational
     temperature and a linewidth of 0.6 km\,s$^{-1}$.}\\
        \tablefoottext{e}{Maximum fractional abundance calculated in the $10^5$-10$^6$ yr range was converted 
    to column density using $N$(H$_2$) = 10$^{22}$ cm$^{-2}$ \citep{CernicharoGuelin1987}.}\\
}
\end{table}

\subsection{Chemical models}
The chemistry of sulfur-bearing molecules in cold dark clouds 
was recently discussed by \cite{Vidal2017}, \cite{Vastel2018}, and \cite{Laas2019}, based on new 
chemical network developments. These studies revealed that the chemistry of sulfur strongly depends on the poorly 
constrained degree of depletion of this element in cold dense clouds. Chemical networks 
are relatively incomplete when dealing with S-bearing species. For example, from the six 
species detected in this work, only C$_4$S is included in the chemical networks 
{\small RATE12} ({\small UMIST}; \citealt{McElroy2013}) and {\small kida.uva.2014} 
({\small KIDA}; \citealt{Wakelam2015}). \cite{Vidal2017} made an effort to expand the number of reactions involving S-bearing species significantly by including several 
of the molecules reported here. These authors, however, discussed only a small number 
of sulfur compounds and did not provide calculated abundances for any of the six 
species reported in this work. We therefore carried out chemical modelling calculations 
to describe the chemistry of the new sulfur-bearing molecules detected. We used the 
gas-phase chemical network {\small RATE12} from the {\small UMIST} database \citep{McElroy2013}, 
expanded with the set of gas-phase reactions involving S-bearing species constructed by 
\cite{Vidal2017}. We included additional reactions to describe the chemistry of NCS and C$_5$S, 
which was not treated by \cite{Vidal2017}, assuming a similar chemical kinetics behaviour to 
NCO and C$_3$S, respectively (see Table~\ref{table:reactions}). Our main purpose is to see whether state-of-the-art gas-phase 
chemical networks can explain the abundances of the S-bearing species discovered. We adopted 
typical parameters of cold dark clouds: a gas kinetic temperature of 10~K, a volume density 
of H$_2$ of $2 \times 10^4$~cm$^{-3}$, a visual extinction of 30~mag, a cosmic-ray ionisation 
rate of H$_2$ of $1.3 \times 10^{-17}$~s$^{-1}$, and 'low-metal' elemental abundances (see, 
e.g. \citealt{Agundez2013}). We therefore adopted a relatively low gas-phase abundance of 
sulfur, $8 \times 10^{-8}$ relative to H.

\begin{figure}
\centering
\includegraphics[angle=0,width=\columnwidth]{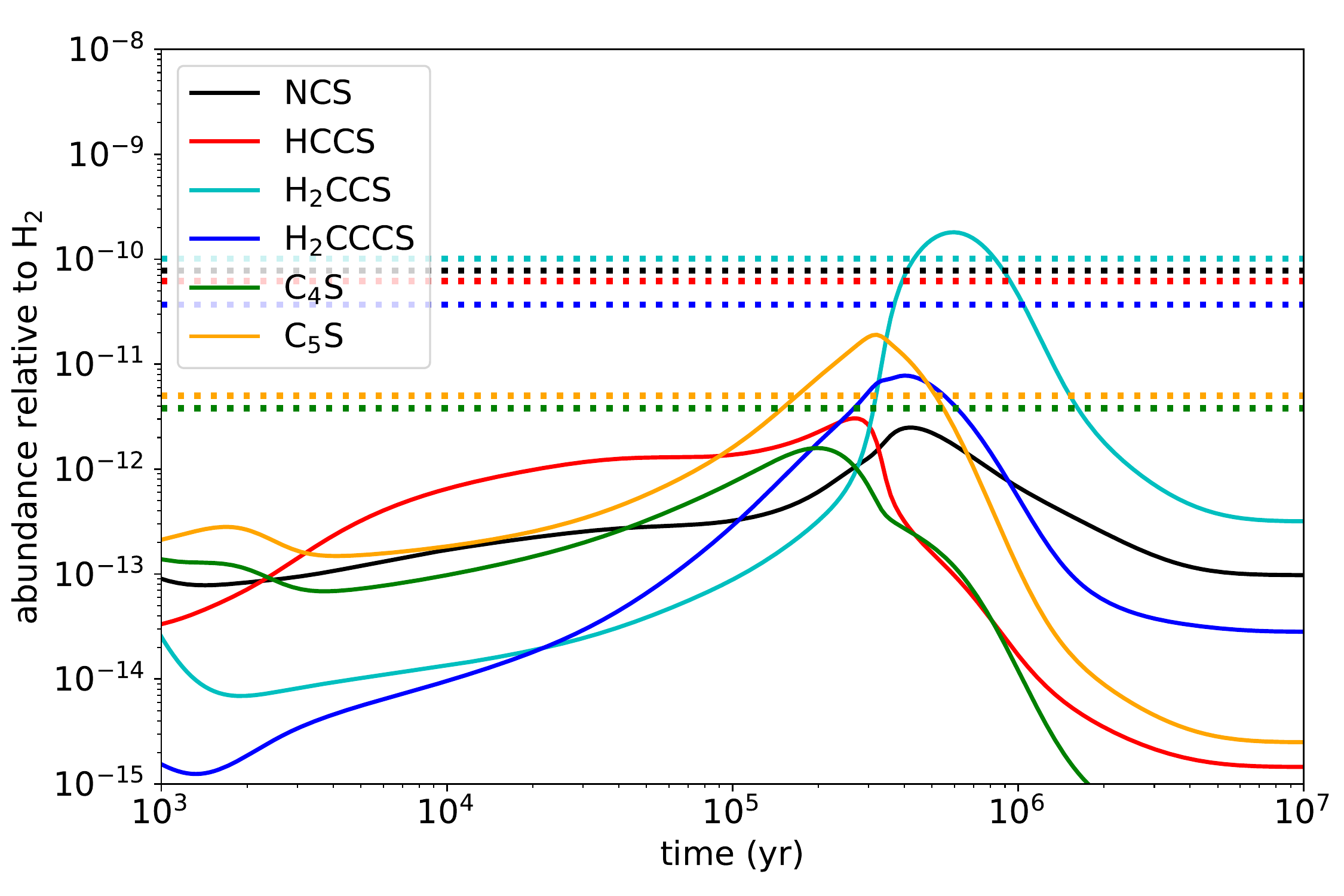}
\caption{Calculated fractional abundances of the six S-bearing species 
reported in this work as a function of time. Horizontal dotted lines correspond to observed 
values in TMC-1 adopting a column density of H$_2$ of 10$^{22}$ cm$^{-2}$ \citep{CernicharoGuelin1987}.} 
\label{fig:abun}
\end{figure}

Calculated column densities for S-bearing species are compared with observed values in 
Table~\ref{column_densities}. In Fig.~\ref{fig:abun} we compare the abundances calculated 
for the six S-bearing species detected in this study with the values derived from the 
observations. The peak abundances calculated are within one order of magnitude of the 
observed abundance for H$_2$CCS, H$_2$CCCS, C$_4$S, and C$_5$S. The two species for 
which the chemical model severely disagrees with observations are NCS and HCCS, in 
which case calculated abundances are lower than observed by more than one order of 
magnitude. These two species have low calculated abundances because they are assumed 
to react quickly with O atoms, with rate coefficients $\gtrsim$\,10$^{-10}$ cm$^3$\,s$^{-1}$. 
This same behaviour has been previously found for other radicals detected in cold dark 
clouds such as HCCO \citep{Agundez2015} and C$_2$S (\citealt{Cernicharo2021a}; see also 
Table~\ref{column_densities}). In these cases, the fast destruction with neutral atoms, 
including O, resulted in calculated abundances well below the observed values. These 
facts suggest that either the abundance of O atoms calculated by gas-phase chemical 
models of cold clouds is too high or O atoms are not as reactive with certain radicals 
as currently thought. While the low-temperature reactivity of O atoms with closed-shell 
S-bearing species such as CS has been shown to be low \citep{Bulut2021}, its reactivity with 
radicals is very poorly known.

The radical NCS is formed in the chemical model through the reactions CN + SO and N + HCS. 
The kinetics and product distribution of these reactions is, however, largely unconstrained. 
This uncertainty also affects the chemical network involving the family of CHNS isomers, for 
which the chemical model underestimates the column densities (see Table~\ref{column_densities}). 
In the chemical models of dark clouds performed by \cite{Adande2010} and \cite{Vidal2017}, the 
main formation pathways to these molecules involve grain surface reactions, which are not 
included in our chemical network.

The formation of the other five molecules reported in this work, which are S-bearing carbon 
chains and thus can be represented by the formula H$_m$C$_n$S, occurs through two types of 
chemical routes. The first one involves neutral-neutral reactions. Along this pathway, HCCS 
is formed by the reactions C + H$_2$CS and OH + C$_3$S, the reaction S + C$_2$H$_3$ yields 
H$_2$CCS, while H$_2$CCCS is mostly formed by the reaction S + CH$_2$CCH. On the other hand, 
the carbon chain C$_4$S is formed through the reactions S + C$_4$H and C + HC$_3$S, while 
C$_5$S is produced in the reactions C$_4$H + CS and S + C$_5$H. It must be noted that the 
kinetics and product distribution of these reactions is very poorly known. Most of these 
reactions are assumed to proceed with capture rate coefficients by \cite{Vidal2017}.

The second pathway consists of reactions involving cations, which ultimately lead to the 
ion H$_p$C$_n$S$^+$, which dissociatively recombines with electrons to yield H$_m$C$_n$S, 
where typically $p = m + 1$. In this route, the ions H$_2$CCS$^+$, H$_3$CCS$^+$, H$_3$CCCS$^+$, 
HC$_4$S$^+$, and HC$_5$S$^+$ are the precursors of HCCS, H$_2$CCS, H$_2$CCCS, C$_4$S, and C$_5$S, 
respectively. The abovementioned ions are in turn formed through reactions between atomic S with hydrocarbon 
ions and S$^+$ with neutral hydrocarbons. In the same line of the reactions discussed above, 
there are large uncertainties regarding these reactions, for which rate coefficients and product 
distributions are taken from \cite{Vidal2017}. The chemical network is probably incomplete in 
that it misses important reactions involving S and S$^+$. Moreover, reactions on grain surfaces, 
which are considered by \cite{Vidal2017} but are not taken into account here, could play an important role.

\section{Conclusions}
In this work, we present the detection of five new sulfur-bearing molecules in
TMC-1: NCS, HCCS, H$_2$CCS, H$_2$CCCS, and C$_4$S. In addition, the species
C$_5$S previously found only towards carbon-rich circumstellar envelopes
is also detected in a cold dark cloud for the first time. Chemical models
fail to reproduce the observed column densities which implies that
the chemical networks are incomplete and that laboratory and theoretical
work has to be performed in order to understand the chemistry of sulfur
in cold prestellar cores.

\begin{acknowledgements}
The Spanish authors thank Ministerio de Ciencia e Innovaci\'on for funding
support through projects AYA2016-75066-C2-1-P, PID2019-106235GB-I00 and
PID2019-107115GB-C21 / AEI / 10.13039/501100011033, and grant RyC-2014-16277. 
We also thank ERC for funding through grant
ERC-2013-Syg-610256-NANOCOSMOS.
\end{acknowledgements}

\begin{appendix}
\section{Column density and rotational temperature determination}
\label{appendix_cd}

In determining the column density of the molecules discovered and analysed in this work,
we adopted a source of uniform brightness and 40$''$ radius \citep{Fosse2001}. We also adopted a 
full linewidth at half power intensity of 0.6 \kms\,, which represents
a good average value to the linewidth of all observed lines in this work.
 
For the species with more than one line observed, we performed rotation 
diagrams to derive
the corresponding rotational temperatures and column densities. 
However, for species for which we observed less than three transitions, we 
adopted the same source parameters and a rotational temperature of 10 K.
In some cases we adopted the rotational temperature 
derived for a similar species.
The different derived and/or adopted rotational temperatures for each
species are given in Table \ref{column_densities} along with the
derived column densities. 
For CS, HCS$^+$, CCS, and C$_3$S, the column densities are taken
from \citet{Cernicharo2021a}. We note that the column
density of CS, as its $J$=1-0 line has a significant optical depth, 
was derived from that of C$^{34}$S adopting the $^{32}$S/$^{34}$S
abundance ratio determined from C$_3$S and C$_3$$^{34}$S (24$\pm$3) \citep{Cernicharo2021a}.
A similar result is obtained from $^{13}$C$^{34}$S adopting a $^{12}$C/$^{13}$C abundance ratio
of 90 \citep{Cernicharo2020c}. For all other molecules studied in this work, the lines
are optically thin.

The error in getting column densities from the observed parameters of a single line
and the assumed rotational temperature has been discussed in detail by \citet{Cernicharo2021a}.
In fact, the main source of uncertainty when the energy of the levels involved in the
transition is $\leq$T$_K$ is not the assumed rotational temperature itself, but the
assumption of it being uniform for all rotational levels \citep{Cernicharo2021a}.
For example, changing the rotational temperature from 5 to 10 K for transitions $J$=1-0 or $J$=2-1 has
a very modest effect on the column density (see, e.g. \citealt{Cernicharo2021a} and \citealt{Cabezas2021}).
This is due to the fact that intensities are proportional to the rotational temperature T$_r$, but the partition function is
inversely proportional to T$_r$ for linear molecules. This also applies to asymmetric species 
with a large rotational constant $A$, which is much larger than T$_r$ (H$_2$CCS and H$_2$CCCS, for example).

As a final check on the derived column densities, we also used the synthetic spectrum resulting from the best fit model 
to all lines of a molecule computed with the MADEX code (Cernicharo 2012). The final parameters 
derived this way should be very similar to
those that could be derived from a standard rotation diagram. Nevertheless, this model fit allows one 
to compare the synthetic line
profiles with observed ones, which is particularly interesting when the rotational transitions exhibit a hyperfine structure as
is the case here for HCS, HSC, and HSCN, and to a lesser extent for HNCS.

\section{Stacking spectra}
\label{stacking}
For C$_4$S and C$_5$S, we observed several individual lines and their detection is solid.
In order to provide additional support for it, we stacked the data following the
procedure used recently to detect CH$_3$CH$_2$CCH in TMC-1 \citep{Cernicharo2021c}.
The data were normalised
to the expected intensity of each transition computed under local thermodynamic equilibrium (LTE) 
for the rotational temperature derived from the
analysis of the individual lines. Finally, all the data were multiplied by the expected 
intensity of the strongest transition among the selected ones. The lines
are optically thin, hence, the intensity of all lines scale in the same way with the assumed column density. Each individual spectra
is weighted as 1/$\sigma_N^2$,
where $\sigma_N$ now accounts for the normalisation intensity factor. 

\begin{figure}[]
\centering
\includegraphics[width=0.89\columnwidth,angle=0]{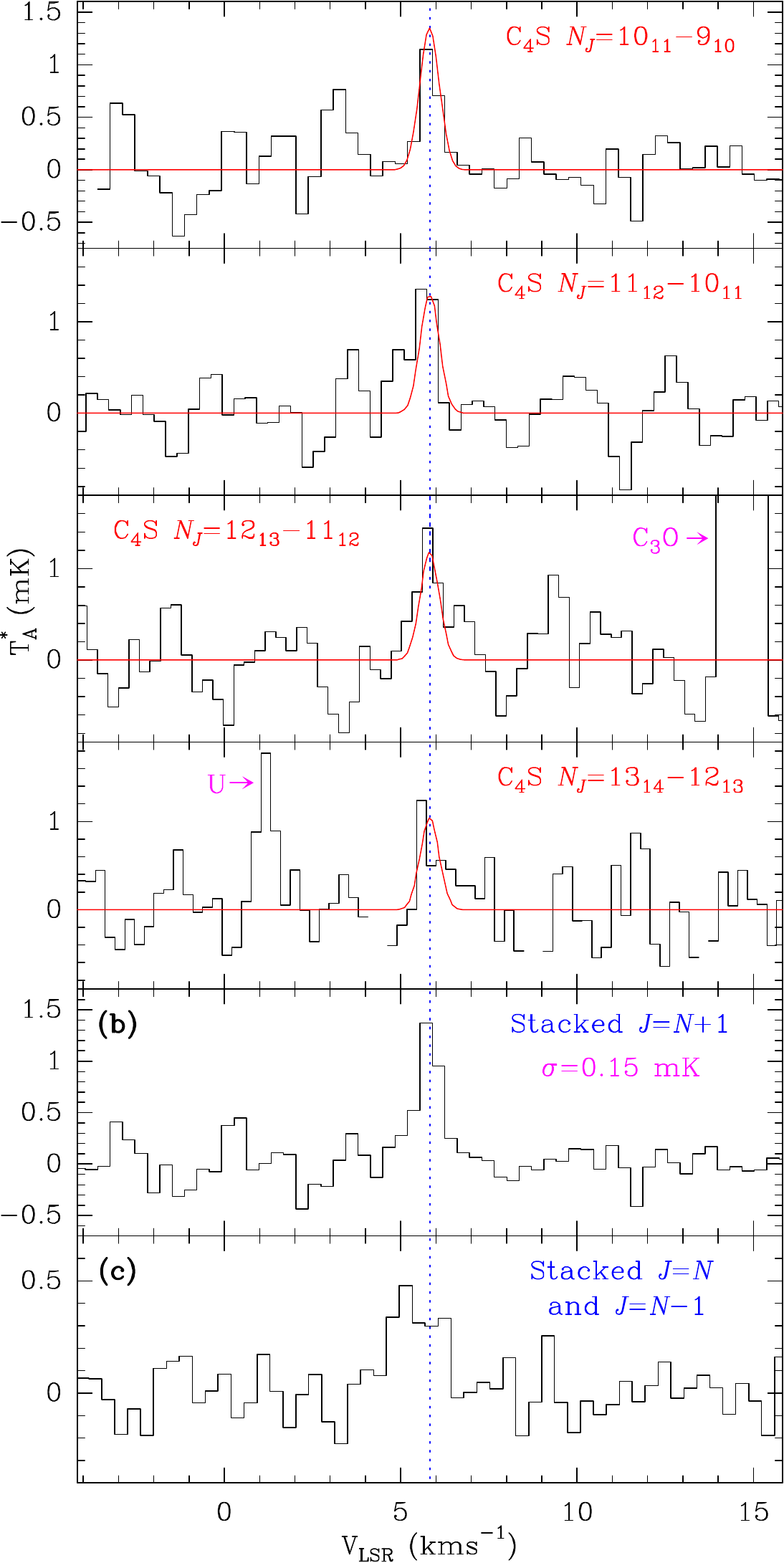}
\caption{
Four upper panels: Individual lines of C$_4$S with $J=N+1$ detected in our survey.
Two bottom panels: Spectrum obtained after stacking all lines of C$_4$S with $J=N+1$
(panel $b$) and the resulting spectrum after stacking the twelve weak lines with $J=N$ and $J=N-1$
(panel $c$). 
%Details on the stacking procedure are provided in Appendix \ref{appendix_cd}.
The abscissa corresponds to the velocity with respect to the local standard of rest.
The dashed blue line corresponds to v$_{LSR}$=5.83 km\,s$^{-1}$.
The ordinate is the antenna temperature in milliKelvin. 
The red line in the four upper panels corresponds to the synthetic spectrum for each transition
computed with T$_r$=7 K and $N$(C$_4$S)=3.8$\times$\ten.}
\label{fig_c4s}
\end{figure}

\begin{figure}[]
\centering
\includegraphics[width=0.89\columnwidth,angle=0]{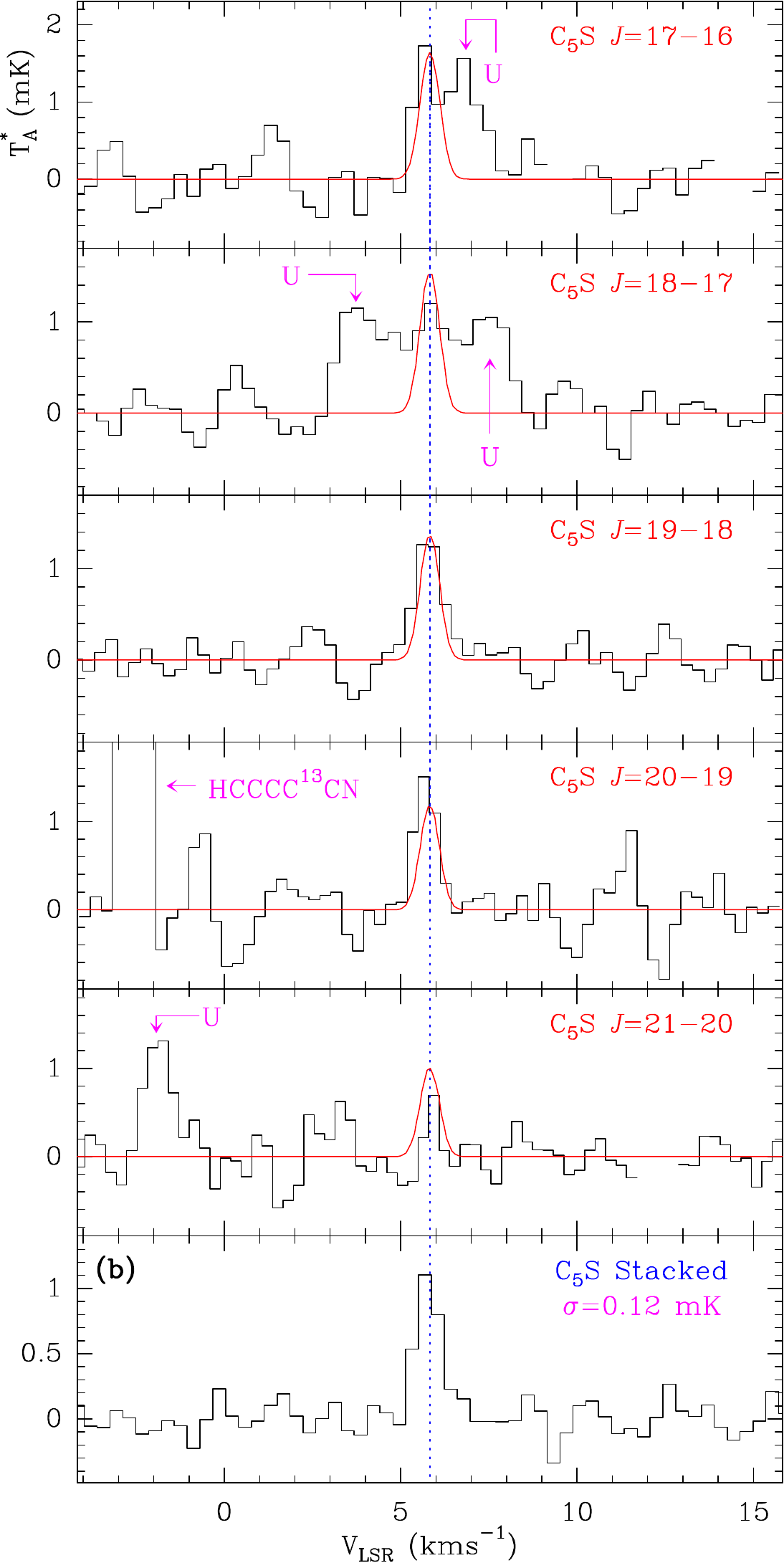}
\caption{Detected lines of C$_5$S within the observed frequency range.
The abscissa corresponds to the velocity with respect to the local standard of rest.
The ordinate is the antenna temperature corrected for atmospheric and telescope losses in milliKelvin. 
Spectral resolution is 38.15 kHz. The dashed blue line corresponds to v$_{LSR}$=5.83 km\,s$^{-1}$.
The red line represents the synthetic spectrum for each line computed for T$_r$=7 K
and $N$(C$_5$S)=5$\times$10$^{10}$ cm$^{-2}$. Panel (b) shows the stacked spectrum
of all rotational transitions of C$_5$S in our survey from $J_u$=17 to 27. 
% See Appendix \ref{appendix_cd} for details).
}
\label{fig_c5s}
\end{figure}

In order to strengthen the detection of C$_4$S,
we grouped the observed lines into two different sets. The first one, with six lines, corresponds to the
strongest transitions for T$_r$=7 K ($J$=$N$+1 transitions).  A second group consists of 
thirteen lines corresponding to the weak $J=N$ and $J=N-1$ transitions, which are 2-3 times weaker
than those of the first group for the derived rotational temperature and, therefore, are below the 3$\sigma$ 
detection limit for individual lines ($\sim$ 0.9-1.5 mK) in the observed frequency range.
We again normalised all spectra to the predicted intensity of the strongest line ($N_J=10_{11}-9_{10}$) 
and averaged all the data in each set using the root mean square noise of each normalised spectrum as a weighting factor. 
We verified that none of these lines are contaminated
by an unknown feature or from a transition of a known molecular species (it could appear centred
on the local standard of rest velocity of the cloud). In all individual spectra,
the emission from other features outside the expected velocity range was removed
by blanking the channels (for example the line of C$_3$O  in the $N_J=12_{13}-11_{12}$ spectrum; see Fig. \ref{fig_c4s}). 
Negative features due to the frequency switching folding were also eliminated before stacking the data.
As quoted above, four of the six
strongest lines are clearly detected (see Fig. \ref{fig_c4s}), and the other two do not show any significant emission at the
expected velocity (their intensities are predicted to be 1.5-2 times weaker than the strongest transition).
The stacked spectrum clearly shows a spectral feature at the
expected velocity with a signal-to-noise ratio of $\simeq$7, as shown in panel $b$ of Fig. \ref{fig_c4s}.
For the group of thirteen weak lines, the resulting stacked spectrum is shown in panel $c$ of the same figure.
A line at 4$\sigma$ is detected. Hence, the detection of the C$_4$S radical is robust.

For C$_5$S, the detection of five individual lines (see Fig. \ref{fig_c5s}) provides solid evidence on the detection of
this species in TMC-1. The observed transitions have upper energy levels between 13 and 20 K. Three
additional transitions above 40 GHz have intensities too low to be detected with the present
sensitivity. Nevertheless, the stacked spectrum of all transitions results on a line at 8$\sigma$. Hence, we conclude
that the detection of these two S-bearing carbon chains in TMC-1 is solid and convincing.

\section{Abundances of related sulfur-bearing species}
\label{s-bearing}
\subsection{HCS and HSC}
\begin{figure}[]
\centering
\includegraphics[width=0.99\columnwidth,angle=0]{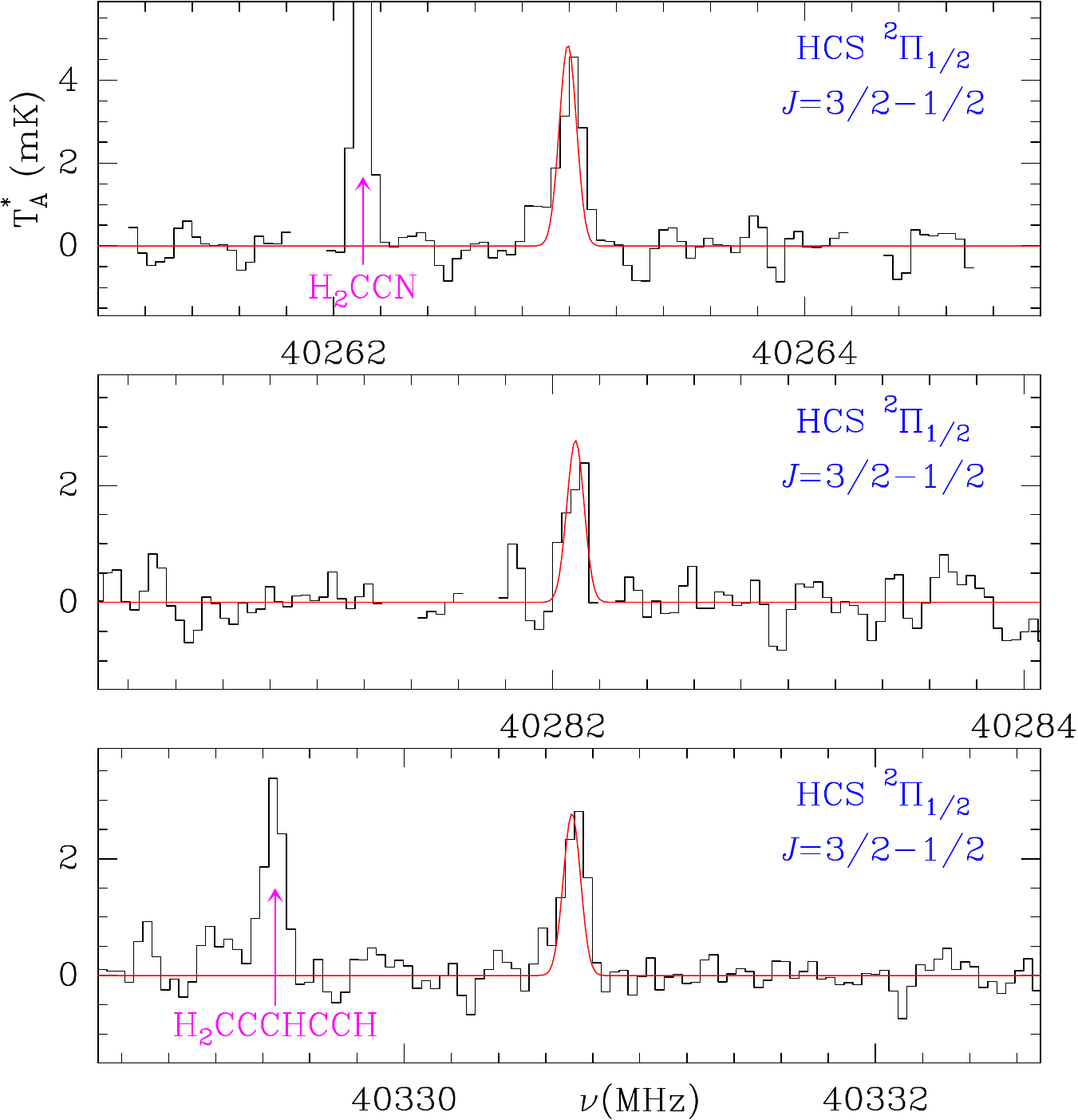}
\caption{Observed lines of HCS within the observed frequency range.
The abscissa corresponds to the rest frequency assuming a velocity with respect to the local standard of rest 
of 5.83 \kms. 
The ordinate is the antenna temperature corrected for atmospheric and telescope losses in milliKelvin. 
Spectral resolution is 38.15 kHz. The red line shows the synthetic spectrum for T$_r$=5 K and 
N=(5.50$\pm$0.50)$\times$10$^{12}$.}
\label{fig_hcs}
\end{figure}

\begin{figure}[]
\centering
\includegraphics[width=0.99\columnwidth,angle=0]{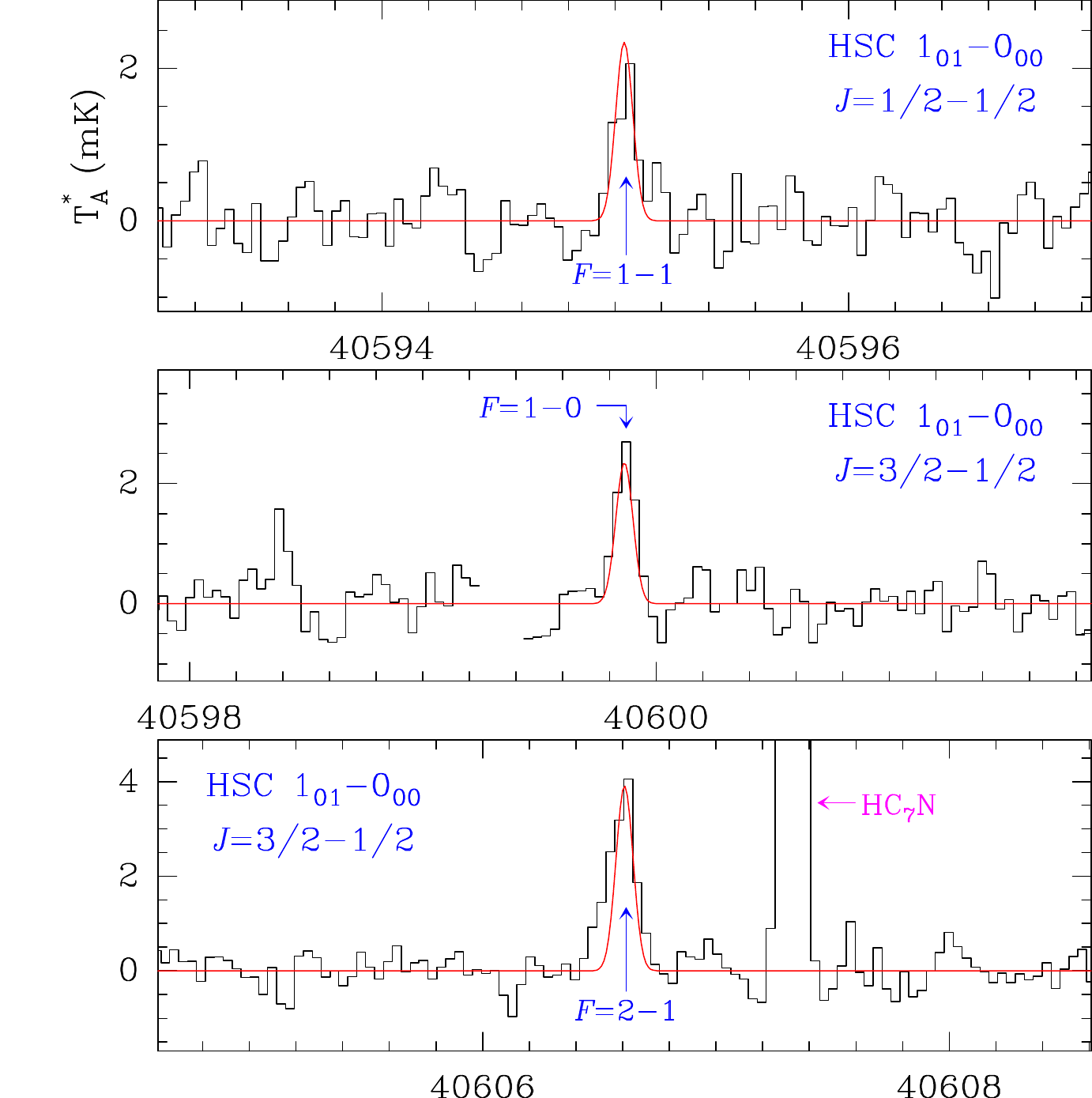}
\caption{Observed lines of HSC within the observed frequency range.
The abscissa corresponds to the rest frequency assuming a velocity with respect to the local standard of rest 
of 5.83 \kms. 
The ordinate is the antenna temperature corrected for atmospheric and telescope losses in milliKelvin. 
Spectral resolution is 38.15 kHz. The red line shows the synthetic spectrum for T$_r$=5 K and 
N=(5.50$\pm$0.50)$\times$10$^{12}$.}
\label{fig_hsc}
\end{figure}

HCS and HSC were discovered by \citet{Agundez2018} towards L483. Here, we
present the first detection of these species in TMC-1. The observed lines are shown in
Figures \ref{fig_hcs} and \ref{fig_hsc}, respectively. Assuming a rotational temperature
of 5 K, we derived $N$(HCS)=(5.5$\pm$0.5)$\times$\doce\  and $N$(HSC)=(1.3$\pm$0.1)$\times$\once.
The column densities increase by $\sim$20\% if a rotational temperature of 10 K is
assumed. The derived abundance ratio $X$(HCS)/$X$(HSC) is 42$\pm$7, which is very
similar to the value derived by \citet{Agundez2018} towards L483. However, we derived
an abundance ratio H$_2$CS/HCS of $\sim$\,9, while in L483 this ratio is $\sim$\,1. 
Our purely gas-phase chemical model results in a H$_2$CS/HCS ratio of $\sim$\,200 
(see Table~\ref{column_densities}) that is too high compared with the values observed in TMC-1 and L483, 
although the chemical model of \citet{Vidal2017}, which includes grain-surface reactions, 
predicts a H$_2$CS/HCS ratio close to the value observed in TMC-1. In the case of HCS and HSC ,
we could speculate with a common precursor, HCSH$^+$.
However, as discussed by \citet{Agundez2018}, the most stable isomer of this
cation is H$_2$CS$^+$ and the formation of HSC from its dissociative recombination
requires a substantial rearrangement of the molecular structure. If, as suggested
by those authors that the CH$_3$S$^+$ cation is the common precursor of H$_2$CS
and HCS, additional reactions are needed to explain the different H$_2$CS/HCS
abundance ratio found towards TMC-1 and L483.

\subsection{SO, H$_2$CS, and OCS}
We note that SO and NS do not have strong transitions in the 31-50 GHz domain; SO has one
transition in this domain, the 3$_2$-2$_2$, but its upper energy level is 21.1 K and we observed a weak emission feature at the correct frequency
(see Table \ref{tab_s_bearing}). This line has peculiar excitation conditions as it
turns out to be in absorption in front of the cosmic microwave background for low volume densities and in
emission for the typical densities of TMC-1. 
\citet{Lique2006} studied SO in detail in this cloud and they derived an abundance of $\sim$10$^{-8}$. Using the collisional rates of 
\citet{Lique2005}, we derived a brightness temperature for the 3$_2$-2$_2$ transition of SO
of -21, -17, 1.8, and 38 mK for n(H$_2$)=10$^4$, 4$\times$10$^4$, 6$\times$10$^4$, and 10$^5$  cm$^{-3}$,
respectively. The predicted intensity has a direct dependency on the column density, but the
turnover between absorption and emission essentially depends on the volume density. From the 
observed parameters for this line (see Table \ref{tab_s_bearing}) 
and the abundance derived by \citet{Lique2006}, we estimate a volume density for the core of 
TMC-1 of 4-6$\times$10$^4$ cm$^{-3}$, which is in good agreement with previous studies (see, e.g. 
\citealt{Fosse2001} and \citealt{Lique2006}). Assuming the H$_2$ column density derived
by \citet{CernicharoGuelin1987}, the column density of SO is $\sim$10$^{14}$ cm$^{-2}$.

%NS has been observed in our survey at 3mm of TMC-1 \citep{Marcelino2007}. The line parameters are
%given in Table \ref{tab_s_bearing}. We have assumed a rotational temperature of 10 K

In our 7mm line survey, H$_2$CS only has one transition. Both, the main and the $^{34}$ isotopologues
have been detected. Line parameters are given in Table \ref{tab_s_bearing}. The column density
for this species was derived assuming a rotational temperature of 10 K \citep{Cernicharo2021a}.

It is important note that OCS has two transitions in the 31-50 GHz domain. Line parameters are given in Table \ref{tab_s_bearing}.

\subsection{HNCS, HSCN, HCNS, and HSNC}
We note that HNCS and HSCN have been previously detected in the interstellar medium \citep{Frerking1979,
Halfen2009}. They are the sulfur equivalent of the well known molecules
HNCO and HOCN; HNCS is the most stable among the possible CHNS isomers, with HSCN, HCSN, and HSNC
being 3200, 17300, and 18100 K above HNCS, respectively \citep{Wierzejewska2003}.

\begin{figure}[]
\centering
\includegraphics[width=0.99\columnwidth,angle=0]{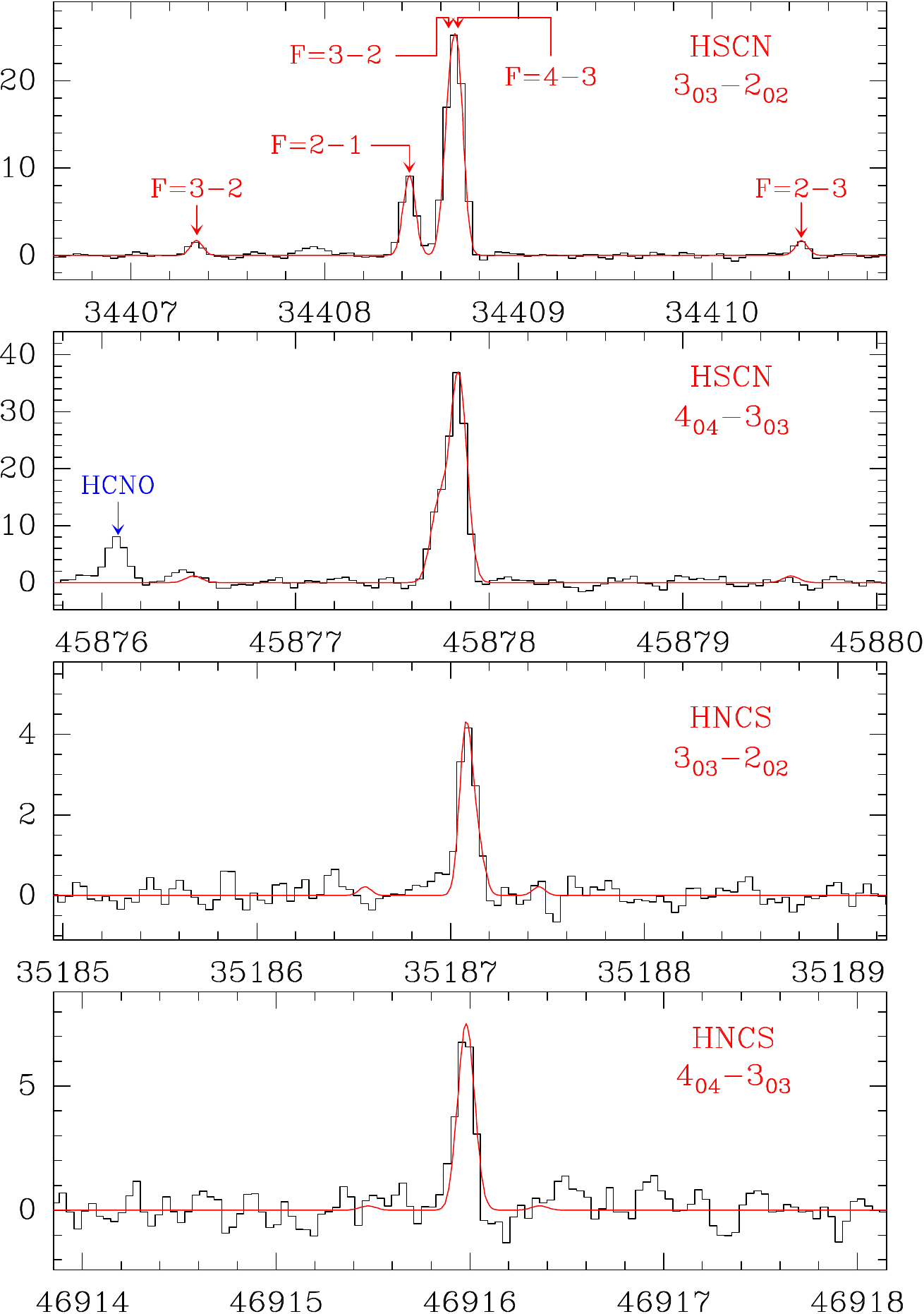}
\caption{Observed lines of HSCN and HNCS within the observed frequency range.
The abscissa corresponds to the rest frequency assuming a velocity with respect to the local standard of rest 
of 5.83 \kms. 
The ordinate is the antenna temperature corrected for atmospheric and telescope losses in milliKelvin. 
Spectral resolution is 38.15 kHz. The red line shows the synthetic spectrum for T$_r$=5 K and 
N=(5.80$\pm$0.60)$\times$10$^{11}$ cm$^{-2}$ for HSCN, and N=(3.80$\pm$0.40)$\times$10$^{11}$ cm$^{-2}$
for HNCS.}
\label{fig_hncs}
\end{figure}

 \citet{Adande2010} have already observed HSCN and HNCS in TMC-1. We observed the 3$_{0,3}$-2$_{0,2}$
and 4$_{0,4}$-3$_{0,3}$ transitions of both 
species, with the corresponding fine structure components for HSCN. The
observed lines are shown in Fig. \ref{fig_hncs}.
By performing a model fitting through the synthetic spectrum, we conclude that
a rotational temperature in the range of 5-8 K produces a good match, although the best
agreement is found for T$_r$=5 K (see Fig. \ref{fig_hncs}). 
We derived a column density of (3.8$\pm$0.4)$\times$\once\, for HNCS and (5.8$\pm$0.6)$\times$\once\, for HSCN. 
It seems that HSCN is slightly more abundant than HNCS, despite being less stable, with an abundance 
ratio of HSCN/HNCS=1.5$\pm$0.3. The column densities of HNCS and HSCN derived by \cite{Adande2010} are 
somewhat lower than those derived here, although the ratio HSCN/HNCS derived by these authors is also around one.

The oxygen analogues HNCO and HOCN have a very different abundance ratio in TMC-1, where 
HNCO/HOCN $\sim$130 \citep{Cernicharo2020c}. The same behaviour is observed in warm molecular clouds, 
where HNCS and HSCN have comparable abundances, while HNCO is two orders of magnitude more abundant 
than HOCN \citep{Adande2010}. These authors argue that HSCN and HCSN could arise from a common precursor, 
HSCNH$^+$, while HNCO may arise mostly from H$_2$NCO$^+$. However, the uncertainties in the chemical 
networks regarding the chemistry of CHNO and CHNS isomers are still large. It is interesting to note 
that while in L483, the abundance ratio HNCO/NCO is $\sim$5 \citep{Marcelino2018}, in TMC-1 the 
HNCS/NCS ratio is $\sim$0.5. This fact strengthens the different behaviour of the CHNO and CHNS 
systems in interstellar clouds.

We also searched for the two isomers HCNS and HSNC (thio and isothio fulminic acid, respectively) 
for which laboratory spectroscopy is available \citep{McGuire2016}. We derived
a 3$\sigma$ upper limit to their column density of 6$\times$\ten\, and 2$\times$\ten, respectively.

\subsection{HCCSH, HCCCSH, and HSCS$^+$}
We note that HCCSH is a metastable isomer, with an energy of 6770 K above the most stable isomer H$_2$CCS
\citep{Lee2019}. It has been searched for towards
several sources \citep{McGuire2019}, but only upper limits have been obtained for its abundance.
In TMC-1, \cite{McGuire2019} derived a 1$\sigma$ upper limit to its column density of $\leq$2.9$\times$\trece.
In our Q-band data, we have searched for the lines of this species. Assuming a rotational
temperature of 7 K, we derived a 3$\sigma$ upper limit of N$\leq$9$\times$\doce.
The rather poor upper limit is due to its low $\mu_a$ dipole moment 
$\mu_a$=0.13\,D \citep{Lee2019}. The $b$ component of the dipole moment is larger, but
$b$-type transitions for this species are outside the frequency range of our survey and they
involve energy levels that will be unpopulated under the physical conditions of TMC-1.

We also searched for HCCCSH, an isomer of H$_2$CCCS, which is predicted to lie around 1000 K above 
it \citep{Brown1982,Crabtree2016}, but only a 3$\sigma$ upper limit to its column density 
of 2.4$\times$\once\, was obtained (see Table \ref{column_densities}). It is interesting to 
note that in TMC-1, the abundance ratio HCCCHO/H$_2$CCCO is well above one 
\citep{Loison2016,Cernicharo2020c}, while the HCCCSH/H$_2$CCCS ratio is found to be below one.

Finally, we have also searched for HSCS$^+$ \citep{McCarthy2009}, the protonated form of carbon disulfide. 
Due to its
low dipole moment \citep{McCarthy2009}, the derived 3$\sigma$ upper limit for its column density is
$\leq$3.0$\times$\doce (average result from the four lines in our survey).

\section{Line parameters}
Line parameters for all observed transitions were derived by fitting a Gaussian line profile to them. A
velocity range of $\pm$20\,\kms\, around each feature was considered for the fit after a polynomial 
baseline was removed. The derived line parameters are
given in Table \ref{tab_s_bearing}.

\clearpage
\onecolumn

%\small
\begin{longtable}{lccccc}
\caption[]{Observed line parameters for sulfur-bearing species in TMC-1.
\label{tab_s_bearing}}\\
\hline
\hline
Transition      & $\nu_{rest}^a$& $\int$T$_A^*$dv $^b$ & v$_{LSR}$$^c$& $\Delta$v$^d$ & T$_A^*$$^e$\\
                &  (MHz)        &  (mK km\,s$^{-1}$)   & (km\,s$^{-1}$)& (km\,s$^{-1}$)& (mK)\\
\hline
\endfirsthead
\caption{continued.}\\
\hline
Transition      & $\nu_{rest}^a$& $\int$T$_A^*$dv $^b$ & v$_{LSR}$$^c$& $\Delta$v$^d$ & T$_A^*$$^e$\\
                &  (MHz)        &  (mK km\,s$^{-1}$)   & (km\,s$^{-1}$)& (km\,s$^{-1}$)& (mK)\\
\hline
\endhead
\hline
\endfoot
\hline
\endlastfoot
\hline
HCCS\\
7/2-5/2 4-3$^*$  & 41077.779$\pm$0.001&   3.6$\pm$0.5& 5.71$\pm$0.02& 0.71$\pm$0.06&    4.8$\pm$0.3\\ 
7/2-5/2 3-2$^*$  & 41078.876$\pm$0.001&   2.3$\pm$0.5& 5.72$\pm$0.03& 0.54$\pm$0.07&    3.4$\pm$0.3\\ 
\hline           
\\
H$_2$CS\\        
$1_{01}-0_{00}$  & 34351.430$\pm$0.001& 413.2$\pm$5.6& 5.85$\pm$0.01& 0.75$\pm$0.01&  518.1$\pm$0.3\\ 
$3_{03}-2_{02}$  &103040.447$\pm$0.001& 577.9$\pm$9.8& 5.80$\pm$0.01& 0.54$\pm$0.01& 1010.7$\pm$2.6\\ 
$3_{13}-2_{12}$  &101477.805$\pm$0.001& 583.4$\pm$9.9& 5.78$\pm$0.01& 0.54$\pm$0.01& 1017.0$\pm$2.5\\ 
$3_{12}-2_{11}$  &104617.027$\pm$0.001& 523.1$\pm$9.5& 5.80$\pm$0.01& 0.54$\pm$0.01&  905.7$\pm$2.2\\ 
\hline           
\\
H$_2$C$^{34}$S\\ 
$1_{01}-0_{00}$  & 33765.750$\pm$0.001&  17.4$\pm$0.8& 5.73$\pm$0.01& 0.77$\pm$0.03&   21.2$\pm$0.4\\ 
$3_{03}-2_{02}$  &101284.314$\pm$0.001&  22.2$\pm$2.0& 5.78$\pm$0.01& 0.52$\pm$0.02&   40.0$\pm$2.4\\
$3_{13}-2_{12}$  & 99774.077$\pm$0.001&  23.8$\pm$1.3& 5.74$\pm$0.01& 0.55$\pm$0.02&   44.0$\pm$1.4\\
$3_{12}-2_{11}$  &102807.337$\pm$0.001&  19.8$\pm$2.8& 5.74$\pm$0.02& 0.53$\pm$0.05&   34.9$\pm$3.0\\
\hline           
\\
H$_2$CCS\\       
$3_{13}-2_{12}$  & 33438.371$\pm$0.003&   1.3$\pm$0.3& 5.91$\pm$0.04& 0.65$\pm$0.08&    1.8$\pm$0.3\\ 
$3_{12}-2_{11}$  & 33783.257$\pm$0.003&   1.9$\pm$0.3& 5.85$\pm$0.04& 0.78$\pm$0.08&    2.3$\pm$0.3\\ 
$4_{14}-3_{13}$  & 44584.335$\pm$0.004&   1.7$\pm$0.3& 5.93$\pm$0.04& 0.55$\pm$0.08&    2.9$\pm$0.4\\ 
$4_{13}-3_{12}$  & 45044.177$\pm$0.004&   2.5$\pm$0.4& 5.87$\pm$0.04& 0.69$\pm$0.09&    3.5$\pm$0.4\\  %blended con línea a 5.33
$7_{17}-6_{16}$  & 78021.269$\pm$0.006&   5.0$\pm$1.0& 5.87$\pm$0.05& 0.32$\pm$0.08&   14.5$\pm$3.8\\
$7_{16}-6_{15}$  & 78825.943$\pm$0.006&              &              &              & $\leq$8.4$^f$ \\
$3_{03}-2_{02}$  & 33611.707$\pm$0.001&   1.5$\pm$0.4& 5.73$\pm$0.07& 1.03$\pm$0.10&    1.4$\pm$0.3\\ 
$4_{04}-3_{03}$  & 44815.323$\pm$0.002&   1.4$\pm$0.4& 5.81$\pm$0.05& 0.67$\pm$0.12&    2.0$\pm$0.3\\ 
\hline           
\\
H$_2$CCCS\\      
$7_{17}-6_{16}$  & 35300.943$\pm$0.007&   3.0$\pm$0.3& 5.78$\pm$0.02& 0.86$\pm$0.05&    3.3$\pm$0.2\\ 
$7_{16}-6_{15}$  & 35466.111$\pm$0.007&   2.3$\pm$0.3& 5.88$\pm$0.03& 0.67$\pm$0.07&    3.3$\pm$0.3\\ 
$8_{18}-7_{17}$  & 40343.856$\pm$0.008&   3.0$\pm$0.4& 5.83$\pm$0.02& 0.60$\pm$0.05&    4.8$\pm$0.3\\ 
$8_{17}-7_{16}$  & 40532.618$\pm$0.008&   3.5$\pm$0.4& 5.92$\pm$0.02& 0.64$\pm$0.04&    5.1$\pm$0.3\\ 
$9_{19}-8_{18}$  & 45386.736$\pm$0.009&   3.9$\pm$0.5& 5.84$\pm$0.03& 0.60$\pm$0.08&    4.6$\pm$0.4\\  %affected by negative feature
$9_{18}-8_{17}$  & 45599.094$\pm$0.009&   3.3$\pm$0.5& 5.97$\pm$0.06& 0.84$\pm$0.18&    3.7$\pm$0.7\\ 
$7_{07}-6_{06}$  & 35384.411$\pm$0.006&   1.0$\pm$0.3& 5.89$\pm$0.05& 0.64$\pm$0.10&    1.4$\pm$0.3\\  %Blended
$8_{08}-7_{07}$  & 40439.230$\pm$0.009&   1.2$\pm$0.3& 5.74$\pm$0.06& 0.62$\pm$0.09&    1.8$\pm$0.3\\ 
$9_{09}-8_{08}$  & 45494.010$\pm$0.009&   1.2$\pm$0.4& 5.92$\pm$0.10& 0.73$\pm$0.13&    1.5$\pm$0.5\\  %Blended
\hline
\\
C$_4$S\\
$10_{11}-9_{10}$ & 32553.160$\pm$0.006&   0.8$\pm$0.3& 5.82$\pm$0.07& 0.63$\pm$0.19&    1.2$\pm$0.3\\ 
$11_{12}-10_{11}$& 35519.754$\pm$0.009&   0.9$\pm$0.4& 5.74$\pm$0.07& 0.51$\pm$0.22&    1.7$\pm$0.3\\ 
$12_{13}-11_{12}$& 38488.028$\pm$0.011&   1.2$\pm$0.4& 5.76$\pm$0.07& 0.85$\pm$0.21&    1.4$\pm$0.3\\ 
$13_{14}-12_{13}$& 41458.049$\pm$0.014&   0.7$\pm$0.3& 5.70$\pm$0.15& 0.45$\pm$0.20&    1.5$\pm$0.3\\ 
\hline
\\
C$_5$S\\
17-16            & 31371.640$\pm$0.010&   0.9$\pm$0.3& 5.65$\pm$0.14& 0.55$\pm$0.15&    1.6$\pm$0.3\\ 
18-17            & 33216.995$\pm$0.012&   0.9$\pm$0.4& 5.78$\pm$0.20& 0.90$\pm$0.25&    1.0$\pm$0.3\\ %BLENDED LEFT AND RIGHT
19-18            & 35062.344$\pm$0.014&   1.1$\pm$0.4& 5.81$\pm$0.07& 0.80$\pm$0.25&    1.2$\pm$0.4\\ 
20-19            & 36907.686$\pm$0.017&   0.7$\pm$0.3& 5.73$\pm$0.10& 0.59$\pm$0.14&    1.1$\pm$0.4\\ 
21-20            & 38753.022$\pm$0.020&   0.4$\pm$0.3& 5.80$\pm$0.10& 0.55$\pm$0.24&    0.8$\pm$0.4\\ 
\hline
\\
HNCS\\
$3_{03}-2_{02}$  & 35187.110$\pm$0.020&   3.7$\pm$0.8& 6.04$\pm$0.03& 0.86$\pm$0.05&    4.0$\pm$0.3\\ 
$4_{04}-3_{03}$  & 46916.000$\pm$0.020&   5.4$\pm$1.0& 6.01$\pm$0.02& 0.70$\pm$0.05&    7.3$\pm$0.6\\ 
$7_{07}-6_{06}$  & 82101.824$\pm$0.002&   5.5$\pm$1.0& 5.89$\pm$0.04& 0.43$\pm$0.08&   12.0$\pm$2.4\\ 
$8_{08}-7_{07}$  & 93830.050$\pm$0.020&   3.5$\pm$1.0& 5.83$\pm$0.05& 0.47$\pm$0.11&    7.1$\pm$1.9\\ 
$9_{09}-7_{07}$  &105558.074$\pm$0.020&   3.7$\pm$1.0& 5.98$\pm$0.04& 0.30$\pm$0.08&   11.7$\pm$3.5\\ 
\hline
\\
HSCN\\
$3_{03}-2_{02}$ 3-2 + 4-3  & 34408.664$\pm$0.003& 25.2$\pm$0.4& 5.86$\pm$0.01& 0.92$\pm$0.01& 25.5$\pm$0.3\\ 
$3_{03}-2_{02}$ 2-1        & 34408.437$\pm$0.003&  7.2$\pm$0.4& 5.95$\pm$0.01& 0.75$\pm$0.03&  9.0$\pm$0.3\\ 
$3_{03}-2_{02}$ 2-2        & 34410.461$\pm$0.003&  1.5$\pm$0.4& 6.00$\pm$0.05& 0.78$\pm$0.12&  1.8$\pm$0.3\\ 
$3_{03}-2_{02}$ 2-2        & 34407.339$\pm$0.003&  0.9$\pm$0.3& 6.00$\pm$0.08& 0.66$\pm$0.14&  1.4$\pm$0.3\\ 
$4_{04}-3_{03}$ 4-3 + 5-4  & 45877.823$\pm$0.004& 24.7$\pm$1.5& 5.80$\pm$0.02& 0.63$\pm$0.03& 36.8$\pm$0.8\\  
$4_{04}-3_{03}$ 3-2        & 45877.736$\pm$0.004&  9.8$\pm$1.6& 5.90$\pm$0.05& 0.68$\pm$0.09& 13.4$\pm$0.8\\  
$7_{07}-6_{06}$            & 80823.192$\pm$0.007& 16.1$\pm$2.5& 5.88$\pm$0.02& 0.57$\pm$0.05& 26.3$\pm$2.7\\ 
$8_{08}-7_{07}$            & 91750.662$\pm$0.008& 13.3$\pm$2.6& 5.95$\pm$0.03& 0.68$\pm$0.07& 18.5$\pm$2.4\\ 
\hline
\\
HCS\\
$1_{01}-0_{00}$ 3/2-1/2 2-1& 40262.996$\pm$0.005&  3.5$\pm$1.0& 5.67$\pm$0.08& 0.76$\pm$0.18&  4.3$\pm$0.3\\ 
$2_{02}-1_{01}$ 5/2-3/2 3-2& 80553.516$\pm$0.004&  6.2$\pm$1.0& 5.64$\pm$0.03& 0.34$\pm$0.07& 17.0$\pm$2.9\\ 
$2_{02}-1_{01}$ 5/2-3/2 2-1& 80565.596$\pm$0.004&  6.2$\pm$1.0& 5.77$\pm$0.07& 0.63$\pm$0.16&  9.3$\pm$2.9\\ 
$2_{02}-1_{01}$ 5/2-3/2 2-1& 80596.409$\pm$0.003&  3.3$\pm$1.0& 5.64$\pm$0.05& 0.28$\pm$0.08& 11.3$\pm$2.7\\ 
\hline
\\
HSC\\
$1_{01}-0_{00}$ 1/2-1/2 0-1& 40584.337$\pm$0.003&  1.2$\pm$0.3& 5.50$\pm$0.06& 0.58$\pm$0.13& 1.9$\pm$0.3\\ 
$1_{01}-0_{00}$ 1/2-1/2 1-1& 40595.040$\pm$0.003&  1.6$\pm$0.4& 5.79$\pm$0.07& 0.88$\pm$0.18& 1.7$\pm$0.3\\ 
$1_{01}-0_{00}$ 1/2-1/2 1-0& 40599.864$\pm$0.003&  2.2$\pm$0.4& 5.81$\pm$0.04& 0.76$\pm$0.09& 2.7$\pm$0.3\\ 
$1_{01}-0_{00}$ 1/2-1/2 2-1& 40606.608$\pm$0.003&  4.0$\pm$0.5& 5.90$\pm$0.04& 1.05$\pm$0.09& 3.6$\pm$0.3\\ 
\hline
\\
OCS\\
4-3                        & 36488.812$\pm$0.000& 25.9$\pm$0.5& 5.88$\pm$0.01& 0.83$\pm$0.01&29.3$\pm$0.3\\
5-4                        & 48651.604$\pm$0.000& 32.1$\pm$0.5& 5.91$\pm$0.01& 0.76$\pm$0.01&39.7$\pm$0.7\\ 
\hline
\\
OC$^{34}$S\\
4-3                        & 35596.869$\pm$0.000&  1.1$\pm$0.4& 5.85$\pm$0.09& 0.92$\pm$0.17& 1.2$\pm$0.3\\
5-4                        & 47762.353$\pm$0.000&             &              &              &$\le$2.1$^f$\\
\hline
\\
SO\\
$3_2-2_2$                  & 36202.034$\pm$0.002&  1.3$\pm$0.0& 6.16$\pm$0.05& 0.70$\pm$0.10& 1.7$\pm$0.3\\
\hline
\end{longtable}
\tablefoot{\\
        \tablefoottext{a}{Predicted frequencies from the MADEX code \citep{Cernicharo2012}. 
    For lines with zero uncertainty for
    the velocity, the frequency corresponds to the measured one assuming v$_{LSR}$5.83km\,s$^{-1}$}.\\
        \tablefoottext{b}{Integrated line intensity in milliKelvin\,km\,s$^{-1}$ }.\\
        \tablefoottext{c}{Line velocity with respect to the local standard of rest in km\,s$^{-1}$. If the
          associated uncertainty is equal to zero, then the velocity has been fixed to 5.83 km\,s$^{-1}$.}\\
        \tablefoottext{d}{Linewidth at half intensity derived by fitting a Gaussian line profile to the observed
     transitions (in km\,s$^{-1}$).}\\
        \tablefoottext{e}{Antenna temperature in milliKelvin.}\\
        \tablefoottext{f}{3$\sigma$ upper limit in mK.}\\
    \tablefoottext{*}{Average of the $e$ and $f$ components which are separated by 33 kHz.}\\    
}
%\normalsize
%\twocolumn

\section{New reactions included}

\begin{table*}
\small
\caption{Reactions involving S-bearing species not included in either the chemical network {\small UMIST RATE12} or that of \cite{Vidal2017}.}
\label{table:reactions}
\centering
\begin{tabular}{lccccl}
\hline \hline
\multicolumn{1}{l}{Reaction} & \multicolumn{1}{c}{$\alpha$ (cm$^3$\,s$^{-1}$)} & \multicolumn{1}{c}{$\beta$} & \multicolumn{1}{c}{$\gamma$ (K)} & \multicolumn{1}{c}{Type} & \multicolumn{1}{l}{Reference} \\
\hline
\multicolumn{6}{c}{NCS chemistry} \\
\hline
CN + S$_2$ $\rightarrow$ NCS + S                            & $2.02 \times 10^{-11}$ &    $-$0.19            & $-$31.9 & 1 & (CN + O$_2$) \\
CN + SO $\rightarrow$ NCS + O                               & $2.02 \times 10^{-11}$ &    $-$0.19            & $-$31.9 & 1 & (CN + O$_2$) \\
N + HCS $\rightarrow$ NCS + H                               & $1.00 \times 10^{-10}$ &     0.00              &       0 & 1 & (N + HCO) \\
O + NCS $\rightarrow$ CO + NS                               & $1.00 \times 10^{-10}$ &     0.00              &       0 & 1 & (O + OCN) \\
H + NCS $\rightarrow$ SH + CN                               & $1.00 \times 10^{-10}$ &     0.00              &       0 & 1 & (H + OCN) \\
C + NCS $\rightarrow$ CS + CN                               & $1.00 \times 10^{-10}$ &     0.00              &       0 & 1 & (C + OCN) \\
H$_3^+$ + NCS $\rightarrow$ HNCS$^+$ + H$_2$                & 0.50                   & $3.47 \times 10^{-9}$ & 2.92    & 2 & (H$_3^+$ + HNCS) \\
H$_3^+$ + NCS $\rightarrow$ HSCN$^+$ + H$_2$                & 0.50                   & $3.47 \times 10^{-9}$ & 2.92    & 2 & (H$_3^+$ + HNCS) \\
HCO$^+$ + NCS $\rightarrow$ HNCS$^+$ + CO                   & 0.50                   & $1.32 \times 10^{-9}$ & 2.92    & 2 & (HCO$^+$ + HNCS) \\
HCO$^+$ + NCS $\rightarrow$ HSCN$^+$ + CO                   & 0.50                   & $1.32 \times 10^{-9}$ & 2.92    & 2 & (HCO$^+$ + HNCS) \\
HNCS$^+$ + H$_2$ $\rightarrow$ H$_2$NCS$^+$ + H             & $1.51 \times 10^{-9}$  &     0.00              &       0 & 1 & (HNCO$^+$ + H$_2$) \\
HNCS$^+$ + H$_2$ $\rightarrow$ HNCSH$^+$ + H                & $1.51 \times 10^{-9}$  &     0.00              &       0 & 1 & (HNCO$^+$ + H$_2$) \\
HSCN$^+$ + H$_2$ $\rightarrow$ H$_2$SCN$^+$ + H             & $1.51 \times 10^{-9}$  &     0.00              &       0 & 1 & (HOCN$^+$ + H$_2$) \\
HSCN$^+$ + H$_2$ $\rightarrow$ HNCSH$^+$ + H                & $1.51 \times 10^{-9}$  &     0.00              &       0 & 1 & (HOCN$^+$ + H$_2$) \\
\hline
\multicolumn{6}{c}{C$_5$S chemistry} \\
\hline
HC$_5$S$^+$ + e$^-$ $\rightarrow$ C$_5$S + H                & $1.00 \times 10^{-7}$  &    $-$0.50            &       0 & 1 & (HC$_3$S$^+$ + e$^-$) \\
HC$_5$S$^+$ + e$^-$ $\rightarrow$ C$_4$H + CS               & $1.00 \times 10^{-7}$  &    $-$0.50            &       0 & 1 & (HC$_3$S$^+$ + e$^-$) \\
HC$_5$S$^+$ + e$^-$ $\rightarrow$ C$_4$S + CH               & $1.00 \times 10^{-7}$  &    $-$0.50            &       0 & 1 & (HC$_3$S$^+$ + e$^-$) \\
C$_4$H + CS $\rightarrow$ C$_5$S + H                        & $2.00 \times 10^{-10}$ &     0.00              &       0 & 1 & (C$_2$H + CS) \\
S + C$_5$H $\rightarrow$ CS + C$_4$H                        & $7.00 \times 10^{-11}$ &     0.00              &       0 & 1 & (S + C$_3$H) \\
S + C$_5$H $\rightarrow$ C$_5$S + H                         & $3.00 \times 10^{-11}$ &     0.00              &       0 & 1 & (S + C$_3$H) \\
C + C$_5$S $\rightarrow$ C$_5$  + CS                        & $3.00 \times 10^{-10}$ &     0.00              &       0 & 1 & (C + C$_3$S) \\
H$_3^+$ + C$_5$S $\rightarrow$ HC$_5$S$^+$ + H$_2$          & 1.00                   & $4.29 \times 10^{-9}$ & 4.39    & 2 & (H$_3^+$ + C$_3$S) \\
H$_3$O$^+$ + C$_5$S $\rightarrow$ HC$_5$S$^+$ + H$_2$O      & 1.00                   & $1.88 \times 10^{-9}$ & 4.39    & 2 & (H$_3$O$^+$ + C$_3$S) \\
HCO$^+$ + C$_5$S $\rightarrow$ HC$_5$S$^+$ + CO             & 1.00                   & $1.61 \times 10^{-9}$ & 4.39    & 2 & (HCO$^+$ + C$_3$S) \\
C$^+$ + C$_5$S $\rightarrow$ C$_5$S+ + C                    & 0.25                   & $2.28 \times 10^{-9}$ & 4.39    & 2 & (C$^+$ + C$_3$S) \\
C$^+$ + C$_5$S $\rightarrow$ C$_6$$^+$ + S                  & 0.25                   & $2.28 \times 10^{-9}$ & 4.39    & 2 & (C$^+$ + C$_3$S) \\
C$^+$ + C$_5$S $\rightarrow$ C$_5$ + + CS                   & 0.25                   & $2.28 \times 10^{-9}$ & 4.39    & 2 & (C$^+$ + C$_3$S) \\
C$^+$ + C$_5$S $\rightarrow$ C$_5$  +CS+                    & 0.25                   & $2.28 \times 10^{-9}$ & 4.39    & 2 & (C$^+$ + C$_3$S) \\
\hline
\multicolumn{6}{c}{Miscellaneous} \\
\hline
H$_3^+$ + H$_2$C$_3$S $\rightarrow$ H$_3$C$_3$S$^+$ + H$_2$ & 1.00                   & $4.10 \times 10^{-9}$ & 2.37    & 2 & KIDA \\
HCO$^+$ + H$_2$C$_3$S $\rightarrow$ H$_3$C$_3$S$^+$ + CO    & 1.00                   & $1.53 \times 10^{-9}$ & 2.37    & 2 & KIDA \\
H$_3$C$_3$S$^+$ + e$^-$ $\rightarrow$ H$_2$C$_3$S + H       & $4.00 \times 10^{-8}$  &    $-$0.70            &       0 & 1 & KIDA \\
H$_3$C$_3$S$^+$ + e$^-$ $\rightarrow$ HC$_3$S + H + H       & $2.00 \times 10^{-7}$  &    $-$0.70            &       0 & 1 & KIDA \\
H$_3$C$_3$S$^+$ + e$^-$ $\rightarrow$ CS + C$_2$H$_2$ + H   & $2.00 \times 10^{-7}$  &    $-$0.70            &       0 & 1 & KIDA \\
H$_3$C$_3$S$^+$ + e$^-$ $\rightarrow$ H$_2$CCC + SH         & $2.00 \times 10^{-7}$  &    $-$0.70            &       0 & 1 & KIDA \\
C$^+$ + H$_2$C$_3$S $\rightarrow$ C$_3$H + HCS$^+$          & 1.00                   & $2.17 \times 10^{-9}$ & 2.37    & 2 & KIDA \\
C + H$_2$C$_3$S $\rightarrow$ CS + C$_3$H$_2$               & $3.00 \times 10^{-10}$ &     0.00              &       0 & 1 & KIDA \\
S + CH$_2$CCH $\rightarrow$ H$_2$C$_3$S + H                & $1.00 \times 10^{-10}$ &     0.00              &       0 & 1 & KIDA \\
S + CH$_2$CCH $\rightarrow$ CS + C$_2$H$_3$                & $2.00 \times 10^{-11}$ &     0.00              &       0 & 1 & KIDA \\
S + CH$_2$CCH $\rightarrow$ HCS + C$_2$H$_2$               & $2.00 \times 10^{-11}$ &     0.00              &       0 & 1 & KIDA \\
H$_3^+$ + HCCS $\rightarrow$ H$_2$C$_2$S$^+$ + H$_2$        & 1.00                   & $2.90 \times 10^{-9}$ & 2.39    & 2 & (H$_3^+$ + H$_2$CCS) \\
HCO$^+$ + HCCS $\rightarrow$ H$_2$C$_2$S$^+$ + CO           & 1.00                   & $1.17 \times 10^{-9}$ & 2.39    & 2 & (HCO$^+$ + H$_2$CCS) \\
 \hline
\end{tabular}
\tablefoot{\\
- For reactions of type 1, the rate coefficient is given by the expression $k(T) = \alpha \, (T/300)^\beta \, \exp(-\gamma/T)$.\\
- For reactions of type 2, the rate coefficient is given by the expression $k(T) = \alpha \, \beta \, (0.62 + 0.4767 \, \gamma \, \sqrt{300/T})$.\\
- Reactions from the KIDA database can be found at \texttt{http://kida.astrophy.u-bordeaux.fr/}.
}
\end{table*}

\end{appendix}

\end{document}